\documentclass{article}

\usepackage{arxiv}

\usepackage[utf8]{inputenc} 
\usepackage[T1]{fontenc}    
\usepackage{hyperref}       
\usepackage{url}            
\usepackage{booktabs}       
\usepackage{amsfonts}       
\usepackage{nicefrac}       
\usepackage{microtype}      
\usepackage{lipsum}
\usepackage{fancyhdr}       
\usepackage{graphicx}       
\usepackage{amsmath}
\usepackage{nomencl}
\graphicspath{{media/}}     

\title{Seis2Rock: A Data-Driven Approach to Direct Petrophysical Inversion of Pre-Stack Seismic Data
}

\author{
  Miguel Corrales \\
  King Abdullah University of Science and Technology (KAUST) \\
  Thuwal, Kingdom of Saudi Arabia\\
  \texttt{miguel.corrales@kaust.edu.sa} \\
   \And
    Hussein Hoteit \\
  King Abdullah University of Science and Technology (KAUST)  \\
  Thuwal, Kingdom of Saudi Arabia\\
  \texttt{hussein.hoteit@kaust.edu.sa} \\
   \And
  Matteo Ravasi \\
  King Abdullah University of Science and Technology (KAUST)  \\
  Thuwal, Kingdom of Saudi Arabia\\
  \texttt{matteo.ravasi@kaust.edu.sa} \\
}

\begin{document}
\maketitle

\begin{abstract}
The inversion of petrophysical parameters from seismic data represents a fundamental step in the process of characterizing the subsurface. We propose a novel, data-driven approach named Seis2Rock that utilizes optimal basis functions learned from well log information to directly link band-limited petrophysical reflectivities to pre-stack seismic data. Seis2Rock is composed of two stages: training and inference. During training, a set of optimal basis functions are identified by performing singular value decomposition on one or more synthetic AVO gathers created from measured or rock-physics synthesized elastic well-logs. In inference, seismic pre-stack data are first projected into a set of band-limited petrophysical properties using the previously computed basis functions; this is followed by regularized post-stack seismic inversion of the individual properties. In this work, we apply the Seis2Rock methodology to a synthetic dataset based on the Smeaheia reservoir model and the open Volve field dataset. Numerical results reveal the ability of the proposed method in recovering accurate porosity, shale content, and water saturation models. Finally, the proposed methodology is applied in the context of reservoir monitoring to invert time-lapse, pre-stack seismic data for water saturation changes.
    
\end{abstract}




\section{Introduction}

Determining petrophysical parameters from seismic data is critical to any hydrocarbon, geothermal, and CO\textsubscript{2} sequestration project. We usually refer to seismic reservoir characterization as the framework under which inversion methods that aim to estimate any form of rock parameters are developed \cite{doyen2007seismic}. Two approaches are commonly used to retrieve petrophysical parameters from pre-stack seismic data. The first, referred to as \textit{sequential or cascaded inversion}, estimates seismic data for elastic parameters; this is followed by a step of rock physics inversion \cite{Mavko2020,Guo2021}. The second approach, named \textit{joint inversion}, aims to estimate elastic and petrophysical parameters simultaneously; this is usually performed in a Bayesian setting using Monte-Carlo sampling methods due to the complex and nonlinear nature of the associated modelling operators \cite{Spikes2007}. In both cases, elastic and petrophysical parameters are linked through empirical relationships retrieved from well logs, core data, laboratory measurements, theoretical rock physics models, or a combination of them. In the context of reservoir modelling, the process of creating a direct link between petrophysical and elastic parameters is usually referred to as Petro-Elastic Modeling (PEM) \cite{Tosaya1982, Smith2003}. On the other hand, pre-stack (or Amplitude Variation with Offset (AVO)) seismic data can be directly modelled from elastic parameters via the nonlinear Zoeppritz equation \cite{zoeppritz1919erdbebenwellen} or by one of its linear approximations \cite{aki1980quantative, fatti1994detection}. Though they are easier to interpret and invert, these linear approximations tend to be valid only for weak parameter contrasts and small angles. \\

The nonlinear relationships linking petrophysical properties to seismic pre-stack amplitudes, in addition to the band-limited nature of seismic data and the inevitable presence of noise, render the seismic to rock parameters a severely ill-posed inverse problem \cite{Bosch2010}. Recent advancements in deep learning have opened new exciting research avenues to handle such nonlinearities. Notable examples of post-stack inversion include the development of a convolutional neural network (CNN) for seismic impedance inversion by \cite{das2019convolutional} and an innovative unsupervised deep-learning method for porosity estimation from post-stack seismic data proposed by \cite{feng2020unsupervised}. Moreover, \cite{smith2022robust} introduced a robust deep learning-based seismic inversion workflow that leverages temporal convolutional networks to transform sequences of post-stack seismic data into a series of predicted acoustic impedance. In the context of pre-stack inversion, \cite{biswas2019prestack} proposed a CNN guided by the physics of the pre-stack (or post-stack) modelling operator, resulting in more accurate and efficient predictions. Likewise, \cite{smith2022robust} utilized two convolutional neural networks to extract petrophysical properties from pre-stack seismic data, further expanding the capabilities of neural network-based inversion methods. The first network, a \textit{direct end-to-end CNN}, emulates the joint inversion approach and exclusively outputs petrophysical properties. In contrast, the second network, a \textit{cascaded CNN}, retrieves both elastic and petrophysical properties as implemented in the conventional sequential approach. \\



Nevertheless, neural networks are well known to be data-hungry, a feature that may not always align with the availability of a limited set of log data (especially in fields with limited well coverage). An alternative data-driven approach to pre-stack seismic inversion was proposed by \cite{Causse2007} under the name of Optimal linear AVO Approximation (OptAVO). This method also uses a priori information in the form of available well-logs and generated seismic reflection curves by means of the nonlinear Zoeppritz equation. Singular Value Decomposition (SVD) is then performed to identify a set of optimal basis functions that are later used to invert pre-stack seismic data into their corresponding elastic parameters. Because a linear relation between elastic parameters and seismic data is found from the data itself, this approach can extract information from seismic amplitudes at a wider angle range than classical model-based approaches. In order to correctly handle wavelet effects in the inversion process, \cite{Ravasi2017} extended the OptAVO approach to band-limited seismic data. Similar to the original approach, explicit inversion of the poorly conditioned AVO operator is circumvented, reducing the impact of noise on the estimated elastic parameters. Compared to the original approach, band-limited OptAVO can retrieve full-bandwidth instead of relative, elastic subsurface models. As an example, \cite{ghaderi2020monitoring} used optimal basis functions to retrieve elastic parameters of the subsurface for estimation of CO\textsubscript{2} saturation using the Sleipner seismic data. \\

In this work, we introduce a data-driven approach to rock physics inversion that extends the capabilities of band-limited OptAVO to perform direct inversion of pre-stack seismic data for petrophysical parameters. We refer to this new approach as \textbf{{Seis2Rock}}. Depending on the availability of elastic (i.e., sonic, shear sonic, and density) well-log data, two variants of Seis2Rock can be identified: first, when elastic well-log data are unavailable, a rock-physics model must be introduced to synthesise such parameters from petrophysical well logs prior to seismic modelling. Second, when elastic well-log data are available, the method becomes fully data-driven in that seismic modelling can be directly performed using the available elastic parameters. In both cases, a training stage is first employed, where a set of elastic properties are used to create a synthetic pre-stack gather employing the Zoeppritz equation. Then, this synthetic gather is used as input to an SVD process, creating a data-driven link between petrophysical parameters and seismic amplitudes. Finally, at the inference stage, petrophysical properties can be estimated by applying the SVD eigenvectors to the pre-stack seismic data and back-projecting the estimated optimal coefficients into petrophysical the parameters of choice (i.e., porosity, shale content, and water saturation). Similar to band-limited OptAVO, the wavelet and time derivative effects are compensate for in an additional step of post-stack seismic inversion. \\



The remainder of this paper is organized as follows. First, we present the theory of the Seis2Rock methodology. This is subsequently validated numerically using a synthetic example created based on Smeaheia's reservoir model. This example is intended to assess the capabilities of our method to retrieve petrophysical properties from pre-stack seismic data, as well as its ability to estimate changes in water saturation in time-lapse settings. Here, we assume that elastic well logs are not available, and therefore perform two steps of modeling in training. A second example is presented based on the field Volve dataset: first, we rigorously describe the sequence of pre-processing steps needed to prepare the data for Seis2Rock. In contrast to the synthetic example, a rock-physics model is not required in this case as the well-log measurements already include elastic-petrophysical pair information, making the method entirely data-driven. The effectiveness of Seis2Rock when dealing with the Volve data is tested by extracting two fences along wells NO 15/9-19 BT2 and NO 15/9-19 A. Initially, the optimal basis functions and coefficients are obtained using well-log information from well NO 15/9-19 BT2. Subsequently, additional well-log information from well NO 15/9-19 A are included in the training process. Finally, we present an analysis and discussion of the results and conclude with a summary of our main findings. \\


\section{The Seis2Rock Method}

\begin{figure}[!h]
    \centering
    \includegraphics[width=0.65\textwidth]{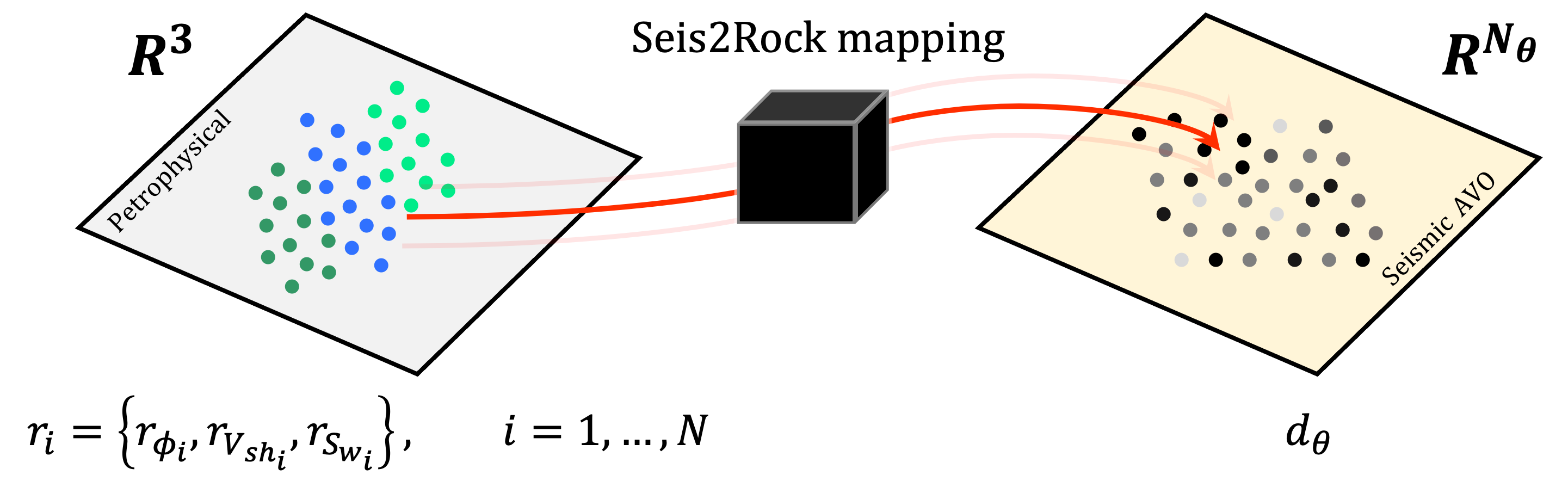}
    \caption{Seis2Rock descriptive goal. Mapping between petrophysical reflectivities and pre-stack seismic amplitudes.}
    \label{fig:RockAVO_mapping}
\end{figure}

Seis2Rock is a data-driven approach to petrophysical inversion \cite{corrales2022data}. Its main goal is to find a direct mapping between physical coefficients (elastic or petrophysical) and seismic AVO responses (Figure \ref{fig:RockAVO_mapping}). Here the physical coefficients are the so-called petrophysical reflectivities ($r_\phi, r_{V_{sh}}, r_{S_w}$), defined as the vertical (time or depth) derivative of the petrophysical properties. Once this direct mapping is found at well locations, the reverse process is applied to all other locations in the seismic data to be inverted.\\

\begin{figure}[!h]
    \centering
    \includegraphics[width=0.95\textwidth]{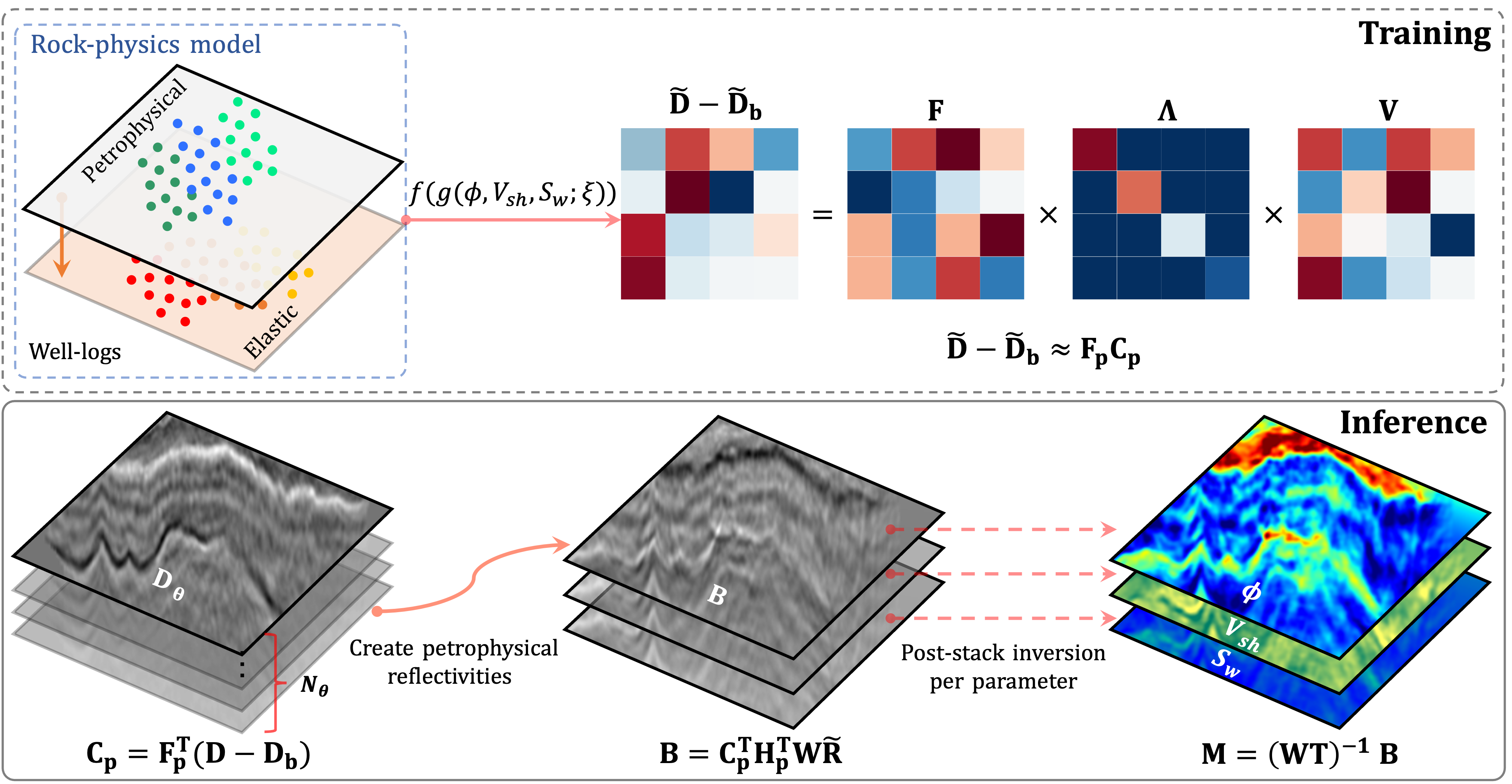}
    \caption{Descriptive summary of Training and Inference stages proposed in Seis2Rock.}
    \label{fig:illustrative_workfow}
\end{figure}

More specifically, the proposed method is composed of two main stages: training and inference. Training refers to the process of obtaining a set of optimal basis functions from pre-stack data modelled at a (small) number of well locations. Inference represents the application of such basis functions to the entire seismic pre-stack data to be inverted. The overall process is summarized in Figure \ref{fig:illustrative_workfow}.\\

\subsection{Training stage}
\begin{figure}[!h]
    \centering
    \includegraphics[width=1\textwidth]{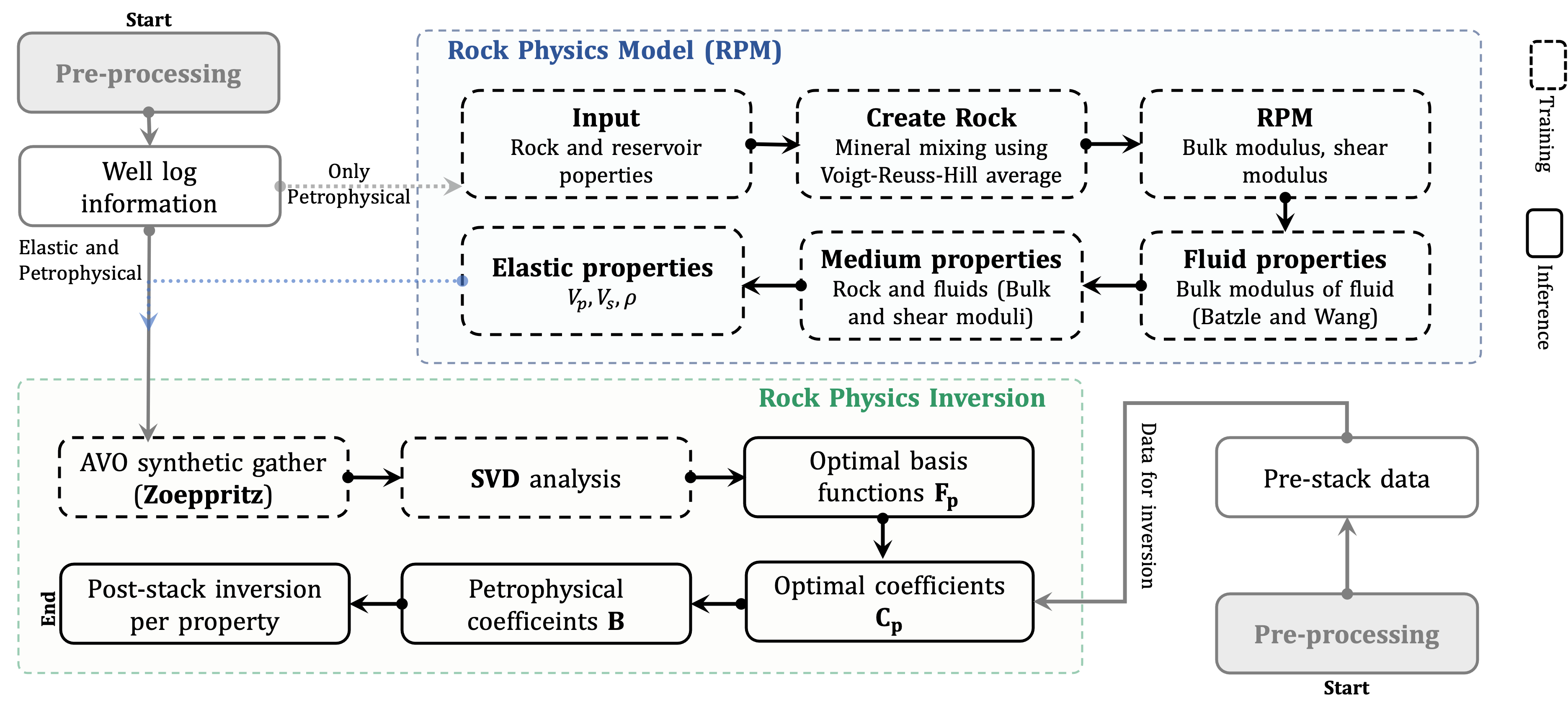}

    \caption{Descriptive workflow of the Seis2Rock methodology, which is composed of training and inference. When the well-log information consists of elastic and petrophysical properties, there is no need to define a Rock-Physics Model.}
    \label{fig:descriptive_workflow}
\end{figure}

In the training stage (see Figure \ref{fig:illustrative_workfow} and \ref{fig:descriptive_workflow}), Seis2Rock aims to find a set of optimal basis functions linking contrasts in petrophysical parameters, also referred to herein as \textit{petrophysical reflectivities}, to pre-stack seismic data. Such petrophysical parameters must be available at one or more well locations as they represent the key training data used by Seis2Rock. If there is a lack of well logs of elastic parameters, the first step of Seis2Rock is represented by the definition of a representative Rock Physics Model (RPM) linking petrophysical and elastic properties; in this work we consider the Hertz-Mindlin model to estimate dry-rock properties such as effective bulk modulus $K_{dry}$ and effective shear modulus $\mu_{dry}$ as a function of pressure $P$, porosity $\phi$, coordination number $C$, shear modulus $\mu_{min}$ and Poisson ratio $\nu_{min}$ of a mixture of minerals.
\begin{equation}\label{kdry}
    \begin{split}
        K_{dry} = \left[  \frac{C^2(1-\phi)^2\mu_{min}^2}{18 \pi^2(1-\nu_{min}^2)} P     \right]^{\frac{1}{3}}\\
    \end{split}
\end{equation} 
\begin{equation}\label{mudry}
    \begin{split}
        \mu_{dry} = \frac{5-4\nu_{min}}{5(2-\nu_{min})}  \left[  \frac{3C^2(1-\phi)^2\mu_{min}^2}{3 \pi^2(1-\nu_{min}^2)} P     \right]^{\frac{1}{3}}\\
    \end{split}
\end{equation} 

Mixing of the minerals (bulk modulus $K_{min}$ and shear modulus $\mu_{min}$) is performed using the Voigt-Reuss-Hill average \cite{Mavko2020}. 
The bulk moduli $K_{fl}$ and density $\rho_{fl}$ of the mixture of fluids are obtained in a similar fashion. Then, Gassmann fluid substitution equations are applied to obtain fluid-saturated moduli and densities:
\begin{equation}\label{Ksat}
    \begin{split}
        K_{sat} &= K_{dry} + \frac{\left( 1-\frac{K_{dry}}{K_{min}} \right)}{\frac{\phi}{K_{fl}}+\frac{1-\phi}{K_{min}}-\frac{K_{dry}}{K_{min}^2}}\\
    \end{split}
\end{equation} 
\begin{equation}\label{musat}
    \begin{split}
        \mu_{sat} &= \mu_{dry}\\
    \end{split}
\end{equation} 
\begin{equation}\label{rho}
    \begin{split}
        \rho &= \rho_{min}(1-\phi) + \rho_{fl}\phi\\
    \end{split}
\end{equation} 

Finally, the wave velocities $V_p$ and $V_s$ are computed as follows: 
\begin{equation}\label{vp}
    \begin{split}
        V_p &= \sqrt{\frac{K_{sat}+\frac{4}{3\mu_{sat}}}{\rho}}\\
    \end{split}
\end{equation} 
\begin{equation}\label{vs}
    \begin{split}
        V_s&= \sqrt{\frac{\mu_{sat}}{\rho}}\\
    \end{split}
\end{equation} 

Equations \ref{kdry} to \ref{vs} represent the chosen nonlinear rock-physics model, compactly defined in the following as $g$. Here porosity ($\phi$), shale content ($V_{sh}$), and water saturation ($S_w$) represent the parameters of the model, while $\xi$ is used to group the set of hyperparameters ($C$, $K$, $\mu$, $\nu$, $P$, and $T$). Finally, given the computed elastic properties ($V_p$, $V_s$, and $\rho$), the synthetic pre-stack seismic data can be obtained via the Zoeppritz equation followed by convolution with the source wavelet. Ultimately, the seismic AVO gather $d$ can be briefly defined as a function of both the nonlinear Zoeppritz equation $f$ and the rock-physics model $g$:
\begin{equation}\label{dgeneral}
    \begin{split}
        d \left(\theta, t \right) = f\left(g (\phi, V_{sh}, S_w; \xi)  \right)\\
    \end{split}
\end{equation} 

Finally, we note that when both elastic and petrophysical information are available in the form of well logs, there is no need to create a rock-physics model, and we can directly apply the \textbf{Rock Physics Inversion} framework illustrated in Figure \ref{fig:descriptive_workflow}.\\

Following the formulation of the problem and the notation used in \cite{Ravasi2017}, SVD is then applied on the synthesised seismic AVO gather to obtain so-called optimal basis functions:
\begin{equation}
    \label{gather}
    \mathbf{\widetilde D} - \mathbf{\widetilde D_b} =
     \mathbf{F} \boldsymbol\Lambda \mathbf{V} = \mathbf{FC}
\end{equation} 

where the matrix $\mathbf{\widetilde D}$ is the modeled seismic gather of size ${N_{\theta} \times N_{t_w}}$ corresponding to the chosen dictionary $\mathbf{\widetilde M}$ of petrophysical parameters of size ${n_m \times N_{t_w}}$. Here $N_{t_w}$ refers to the size of the vertical axis used for both the well logs and seismic data and $N_{\theta}$ is the number of angles. Finally, $n_m$ refers to the number of independent petrophysical parameters used in the rock physics model. Similarly, $\mathbf{\widetilde D_b}$ represents the seismic AVO gather modelled which is also used to obtain the background model $\mathbf{\widetilde M_b}$ (i.e., a smoothed version of the petrophysical parameters $\mathbf{\widetilde M}$). Singular vectors are placed in matrices $\mathbf{F}$ and $\mathbf{V}$ of size $N_{\theta} \times N_{t_w}$ and $N_{t_w} \times N_{t_w}$, respectively, whilst singular values are placed along the diagonal of the matrix $\boldsymbol\Lambda$ of size $N_{t_w} \times N_{t_w}$. Matrices $\boldsymbol\Lambda$ and $\mathbf{V}$ are conveniently combined to form the matrix of optimal coefficients $\mathbf{C}$, which act as the weights of the basis functions stored along each column of the matrix $\mathbf{F}$ to form the data $\mathbf{\widetilde D} - \mathbf{\widetilde D_b}$. When the SVD process is applied to the modeled seismic data, the singular values tend to quickly decay to zero, meaning that the contribution of the different basis functions to the reconstruction of the data rapidly decreases; therefore it is possible to consider a small subset of basis functions $p<N_{\theta}$, and write an approximate relation as follows:
\begin{equation}
    \label{gather_p}
    \begin{split}
        \mathbf{\widetilde D} - \mathbf{\widetilde D_b} \approx \mathbf{F_p} \mathbf{C_p}\\
    \end{split}
\end{equation} 

where $\mathbf{F_p}$ and $\mathbf{C_p}$ are matrices of size $N_{\theta} \times p$ and $p \times N_{t_w}$, respectively. Equation \ref{gather_p} marks the end of the training process and provides the optimal basis functions $\mathbf{F_p}$ for the inference process. Although we have considered here petrophysical logs for a single well, Seis2Rock can be easily extended to accommodate for the availability of multiple wells. This can be accomplished by concatenating $n$ well profiles one after the other whilst ensuring a smooth transition in the properties between two consecutive wells. As a consequence, the sizes of the corresponding data $\mathbf{\widetilde D}$ and dictionary $\mathbf{\widetilde M}$ matrices become ${N_{\theta} \times n \cdot N_{t_w}}$ and ${n_m \times n \cdot N_{t_w}}$.

\subsection{Inference stage}
The inference process intends to convert a seismic pre-stack dataset covering an extensive geographical area distant to the control well(s) into a set of petrophysical property models.
Here we consider for simplicity a single location, and define
$\textbf{D}$ and $\textbf{M}$ to be matrices of size $N_{\theta} \times N_t$  of size $n_m \times N_t$, respectively. Since $\mathbf{F_p}$ is an orthonormal matrix by construction, the optimal coefficients for any seismic gather $\textbf{D}$ (i.e., at a given spatial location) can be obtained as follows:
\begin{equation}\label{Cp}
    \begin{split}
        \mathbf{C_p}= \mathbf{F^T_p} (\mathbf{D}-\mathbf{D_b})
    \end{split}
\end{equation} 

Similarly to the training process, a consistent background model $\mathbf{M_b}$ is required to model a background synthetic dataset $\mathbf{D_b}$ to be subtracted from the recorded data. These optimal coefficients are subsequently back-projected into a band-limited representation of the physical petrophysical parameters $\mathbf{B}$ (see Appendix \ref{appendix_A} for the full derivation of the back-projection process). Finally, an inverse problem is solved to undo the effect of the wavelet $\mathbf{W}$ and time-derivative $\mathbf{T}$ operators:
\begin{equation}\label{m}
    \begin{split}
        \mathbf{M}= (\mathbf{WT})^{-1} \mathbf{C_p^T} \mathbf{H_p^T} \mathbf{W} \widetilde{\mathbf{R}} = (\mathbf{WT})^{-1}\mathbf{B}
    \end{split}
\end{equation} 

where $\mathbf{\widetilde{R}}$ is the matrix containing the petrophysical reflectivities from the well log used in training of size $n\cdot N_t \times n_m$, and $\mathbf{H_p}=\mathbf{V_p^T} \boldsymbol\Lambda_\mathbf{p}^{-1}$ is a matrix of size $N_t \times p$. Note that the right-hand-side of equation \ref{m} can be interpreted as a series of post-stack seismic inversions (one post-stack inversion per petrophysical parameter), which are solved here using the PyLops computational framework \cite{Ravasi2020}. Moreover, whilst we have considered a single location for simplicity in this derivation, the final step of inversion is usually carried out for all spatial locations at the same time, such that spatial regularization in the form of Laplacian or Total Variation (e.g., \cite{ravasi2022}) can be introduced.

\section{Results}
\subsection{Synthetic data}
\begin{figure}[!h]
    \centering    \includegraphics[width=\textwidth]{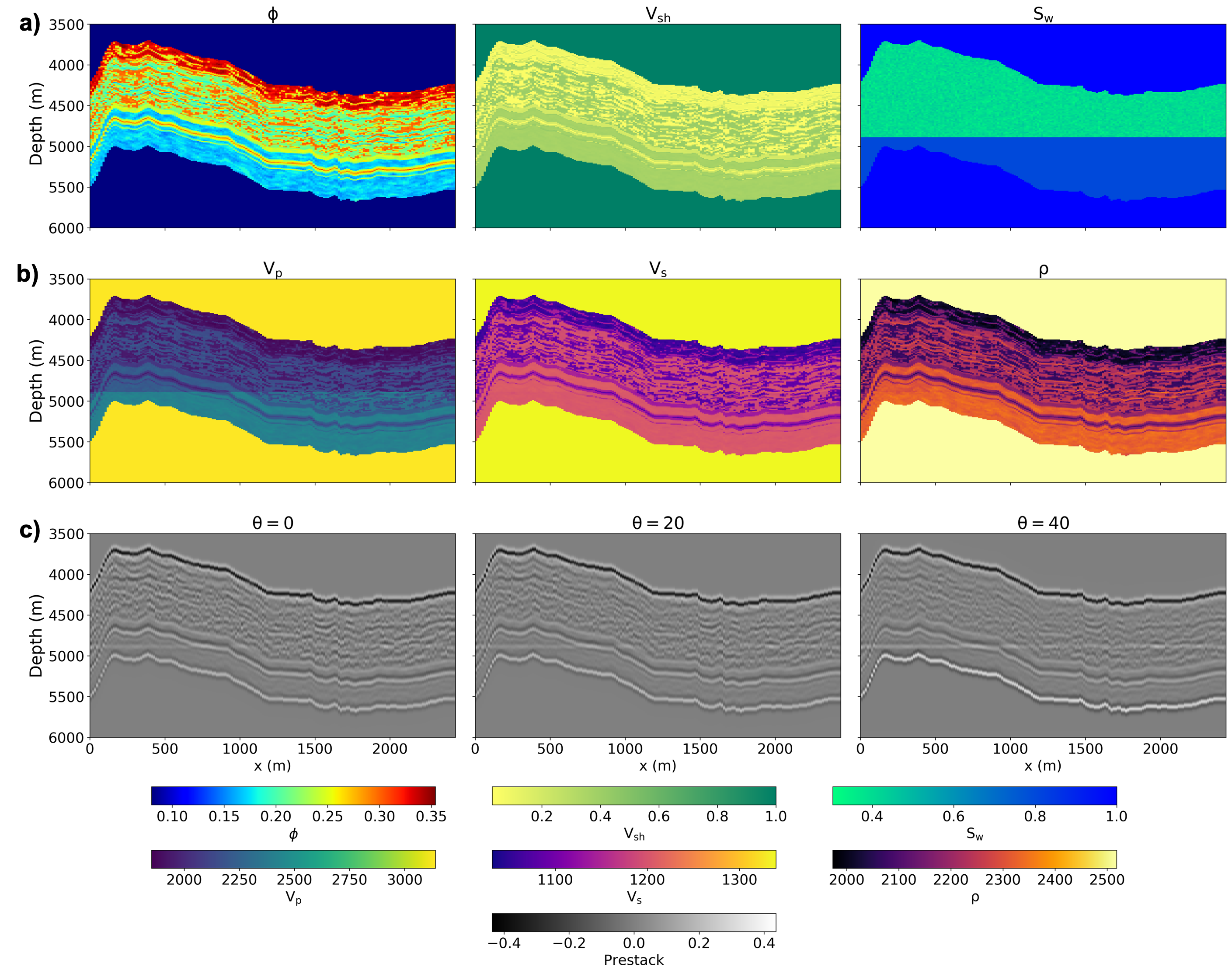}
    \caption{Synthetic 2D example based on Smeheia Reservoir model. a) Petrophysical properties. b) Derived Elastic properties using the RPM. c) AVO synthetic gather using Zoeppritz equation.}
    \label{fig:setting_synthetic}
\end{figure}

The proposed method is first assessed on a synthetic example. The porosity model is constructed based on a 2D section of the Smeaheia reservoir model, whilst the shale content model is stochastically generated using a normal-random distribution conditioned to the porosity values. Likewise, the water saturation model is assumed to follow a normal random distribution above the water contact (Depth=4895 m). The system is assumed to be only occupied by oil and water and the reservoir conditions are assumed to be a pressure of 24.1 MPa, temperature of 50 C, water salinity of 10,000 ppm, and oil gravity of 20 API. Figure \ref{fig:setting_synthetic}a shows the true petrophysical models. \\


\newpage

\subsubsection{Direct Petrophysical Inversion}

In this case, different vertical pillars in the model are assumed to correspond to wells. Based on the information in the selected well logs ($\phi, V_{sh}, S_w$) and the hyperparameters $\xi=$ $\{K_{sand}= 37.6 \times 10^9$ Pa, $\mu_{sand}= 44.6 \times 10^9$ Pa,  $\rho_{sand}= 2.65$ g/cm$^3$; and $K_{shale}= 20.9 \times 10^9$ Pa, $\mu_{shale}= 30.6 \times 10^9$ Pa, and  $\rho_{shale}= 2.58$ g/cm$^3 \}$, the rock-physics model in equations \ref{kdry}-\ref{vs} is used to compute the elastic parameters in the different wells and the entire 2D sections \ref{fig:setting_synthetic}b. Such elastic parameters are further used to model reflection coefficients for angles ranging from $0^\circ$ to $50^\circ$ \ref{fig:setting_synthetic}c. Whilst angles beyond $30^\circ$ cannot be handled by conventional linear approximations, this example is created to prove that our methodology can handle such angles and is therefore more successful in recovering strong contrasts. Finally, the reflection coefficients are band-passed using a Ricker wavelet with a central frequency of 20 Hz to produce the seismic AVO gathers, which are used as input to the SVD process as well as the data to be inverted. \\

In the following setup, three experiments are performed assuming knowledge of petrophysical well-logs at one, two, or three vertical profiles ($x = 281.6, 1182.8, 1971.32 ~ \mathrm{m}$). By including the wells progressively, we wish to analyze the impact on the final inversion results of having more information in the training process. Regardless of on the amount of wells available in the training process, $p=6$ coefficients are chosen to decompose the data, as in equation \ref{gather_p}. Using equation \ref{Cp}, the retrieved optimal basis functions are used to estimate the band-limited coefficients. Finally, to obtain the petrophysical parameters, each band-limited petrophysical reflectivity model is inverted by means of spatially regularized post-stack inversion where a Laplacian operator is used as a regularizer. Furthermore, we do not impose any iterative constraint on the inversion within the range [0,1] for each output produced by the petrophysical inversion process. \\

\begin{figure}[!h]
    \centering
    \includegraphics[width=\textwidth]{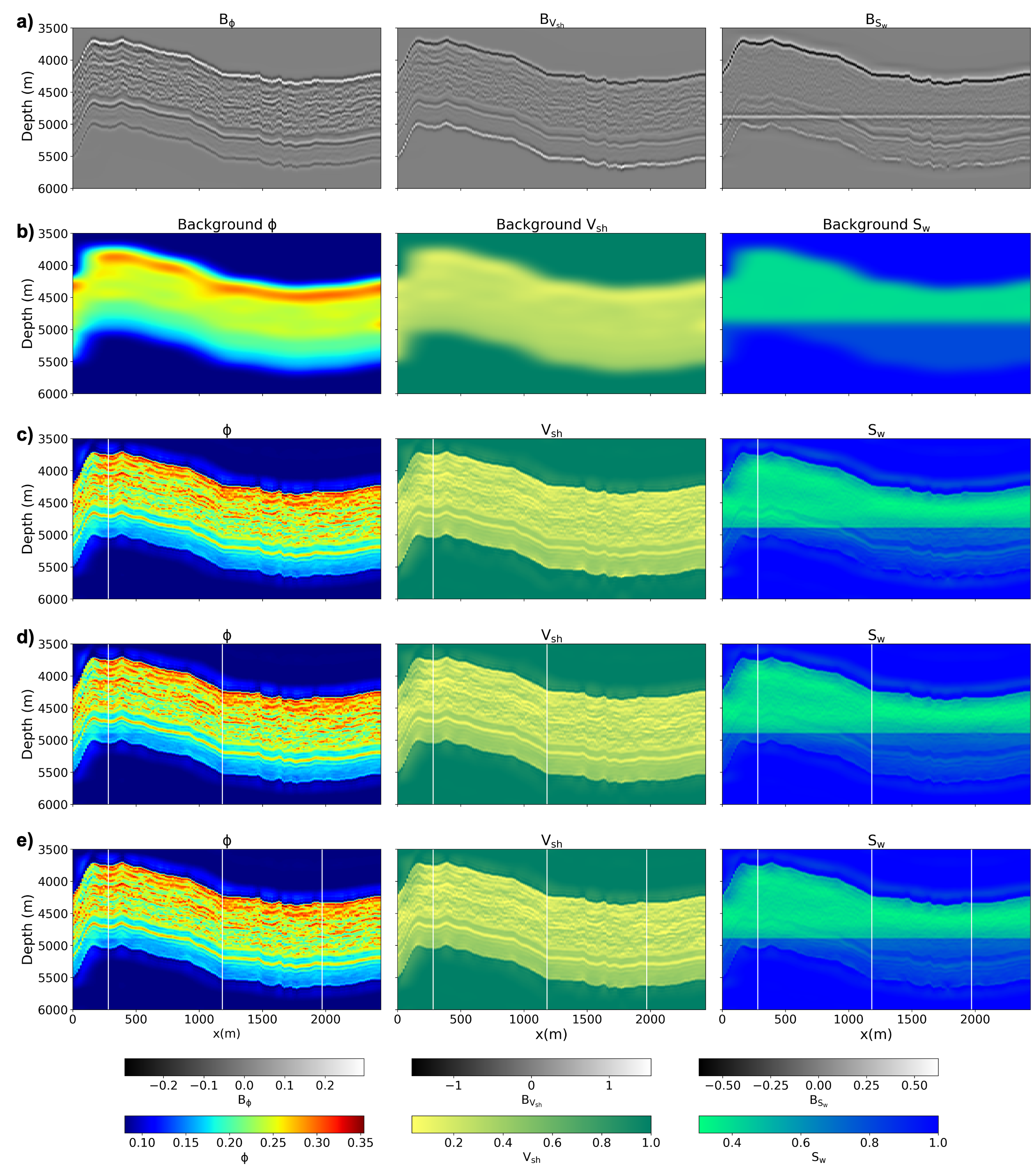}
    \caption{Results of inversion stacking  wells during the training.  a) Example of Petrophysical coefficients obtained when using one well to build the optimal basis functions. b) Set of background models used for inversion. Panels c), d), and e) show the inversion results when adding one, two, and three wells on the training stage respectively. The vertical white lines represent the location of the well logs extracted to build the optimal basis functions.}
    \label{fig:results_synthetic_stacking}
\end{figure}

\begin{figure}[h]
    \centering
    \includegraphics[width=\textwidth]{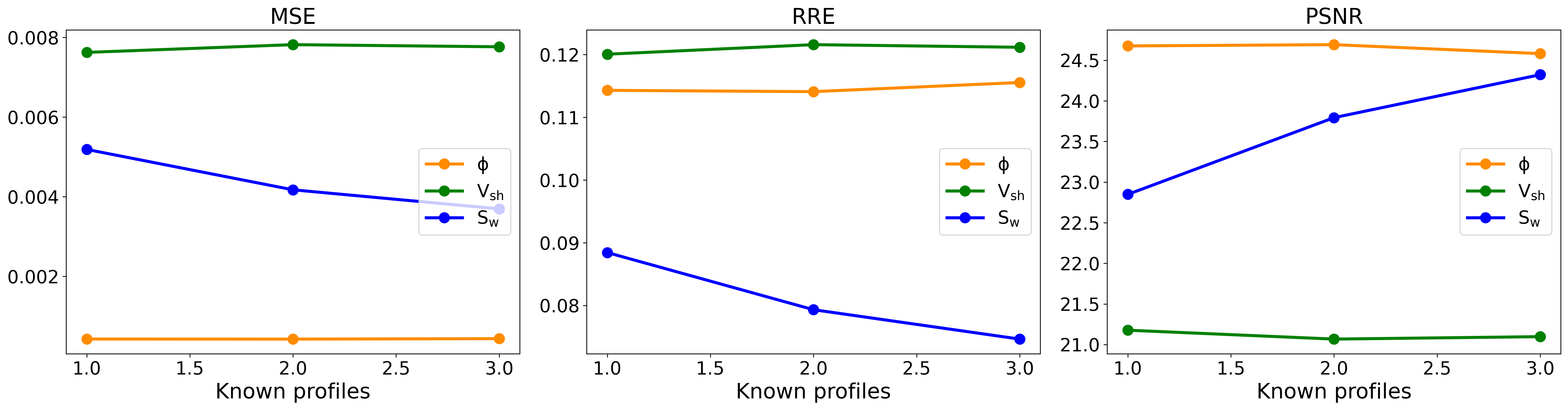}
    \caption{Metrics summarizing the quality of the inversion process as the number of wells used to build the basis functions increases. From left to right, mean square error (MSE), relative residual error (RRE), and peak signal-to-noise ratio (PSNR). Water saturation improves while adding more wells for training in this example.}
    \label{fig:errors_metrics}
\end{figure}

\begin{figure}[h]
    \centering
    \includegraphics[width=\textwidth]{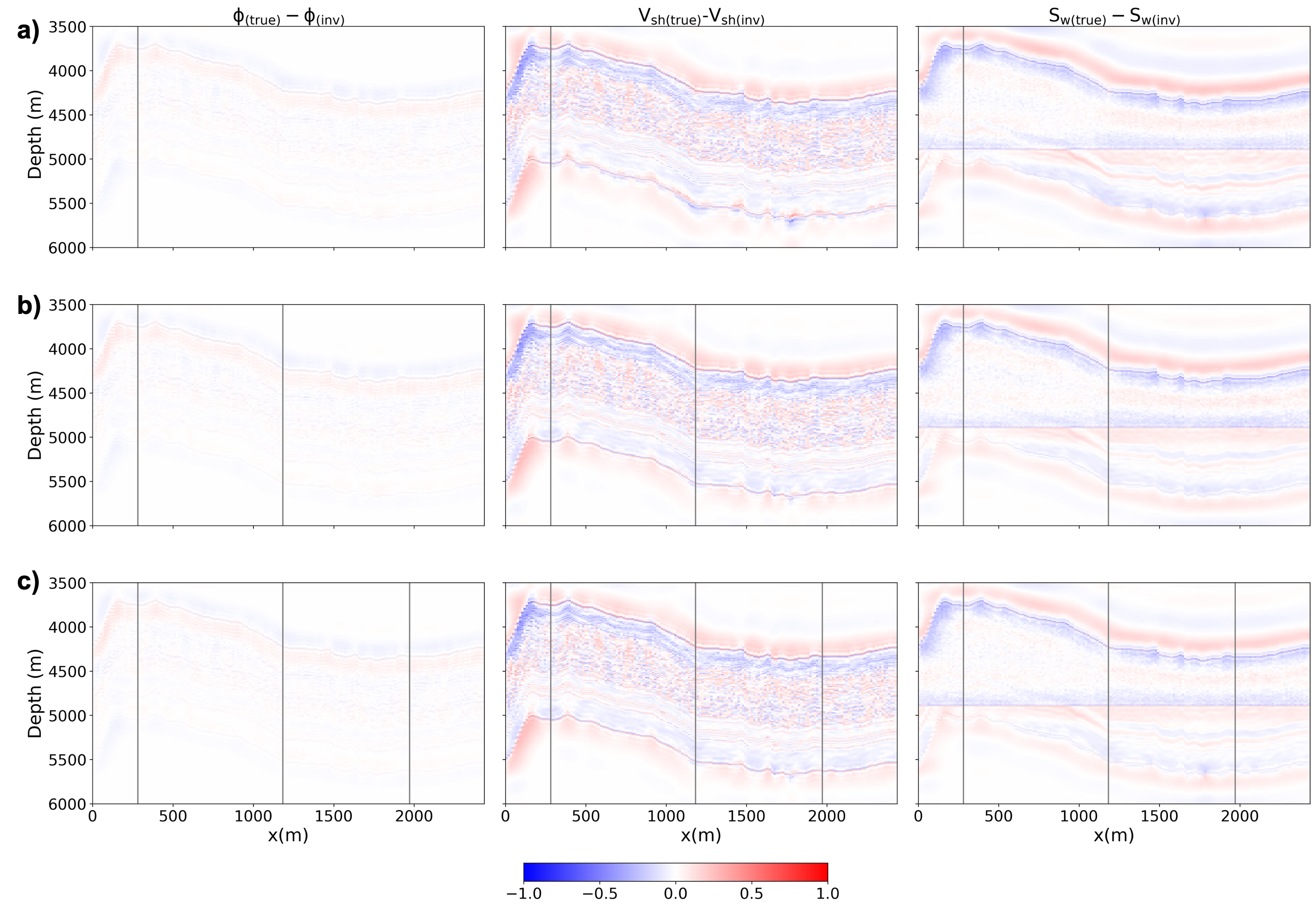}
    \caption{Difference between the true and the inverted petrophysical models when using a) one, b) two, and c) three wells in the training stage respectively.}
    \label{fig:results_synthetic_abs_error}
\end{figure}

Figure \ref{fig:results_synthetic_stacking} shows the inverted petrophysical parameters for the cases with one, two, and three vertical profiles, exhibiting high accuracy in all cases. In addition, the absolute error is visualized in Figure \ref{fig:results_synthetic_abs_error} to further assess the quality of the reconstruction process. Moreover, the mean square error (MSE), relative residual error (RRE), and peak signal-to-noise ratio (PSNR) of the inverted parameters are computed as a function of the number of wells included in the training process, showing an increment in the quality of the inversion as more information is added to the training process (Figure \ref{fig:errors_metrics}). From this figure, we can also observe that porosity is the best-resolved parameter, followed by water saturation and shale content; this is a direct consequence of the fact that elastic parameters (and therefore seismic data) are more sensitive to porosity and saturation variations in the pores than rock type variations in the matrix. \\

\begin{figure}[h]
    \centering
    \includegraphics[width=\textwidth]{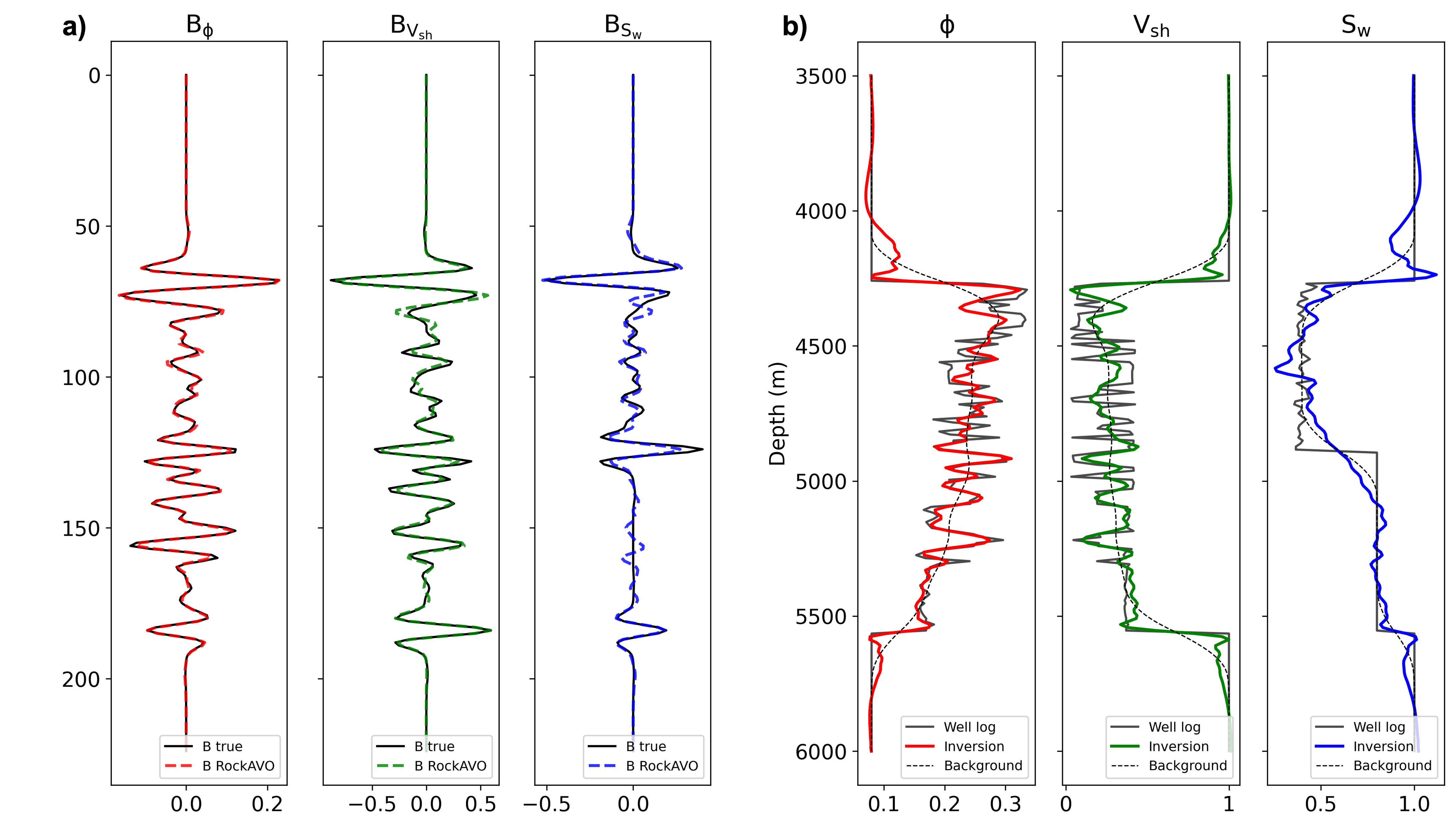}
    \caption{Inversion results along well $x=1971.3 ~\mathrm{m}$. a) Band-limited petrophysical coefficients (p=6) and b) 
 results of petrophysical inversion.}
    \label{fig:bcoeff_and_well_inversion}
\end{figure}

In addition, the inversion results along the well at location x=1971.3 m are presented in Figure \ref{fig:bcoeff_and_well_inversion}b. The recovered properties are highly accurate when compared to the ground truth, except in places where high contrasts are present; this issue could be circumvented using another type of regularization, as explained in more detail in the Discussion section. Finally, Figure \ref{fig:bcoeff_and_well_inversion}a presents a comparison between the true petrophysical coefficients and the petrophysical coefficients obtained using only the first p=6 coefficients (or singular values). \\


\clearpage

\subsubsection{Seis2Rock as water saturation tracker}
\begin{figure}[!h]
    \centering
    \includegraphics[width=\textwidth]{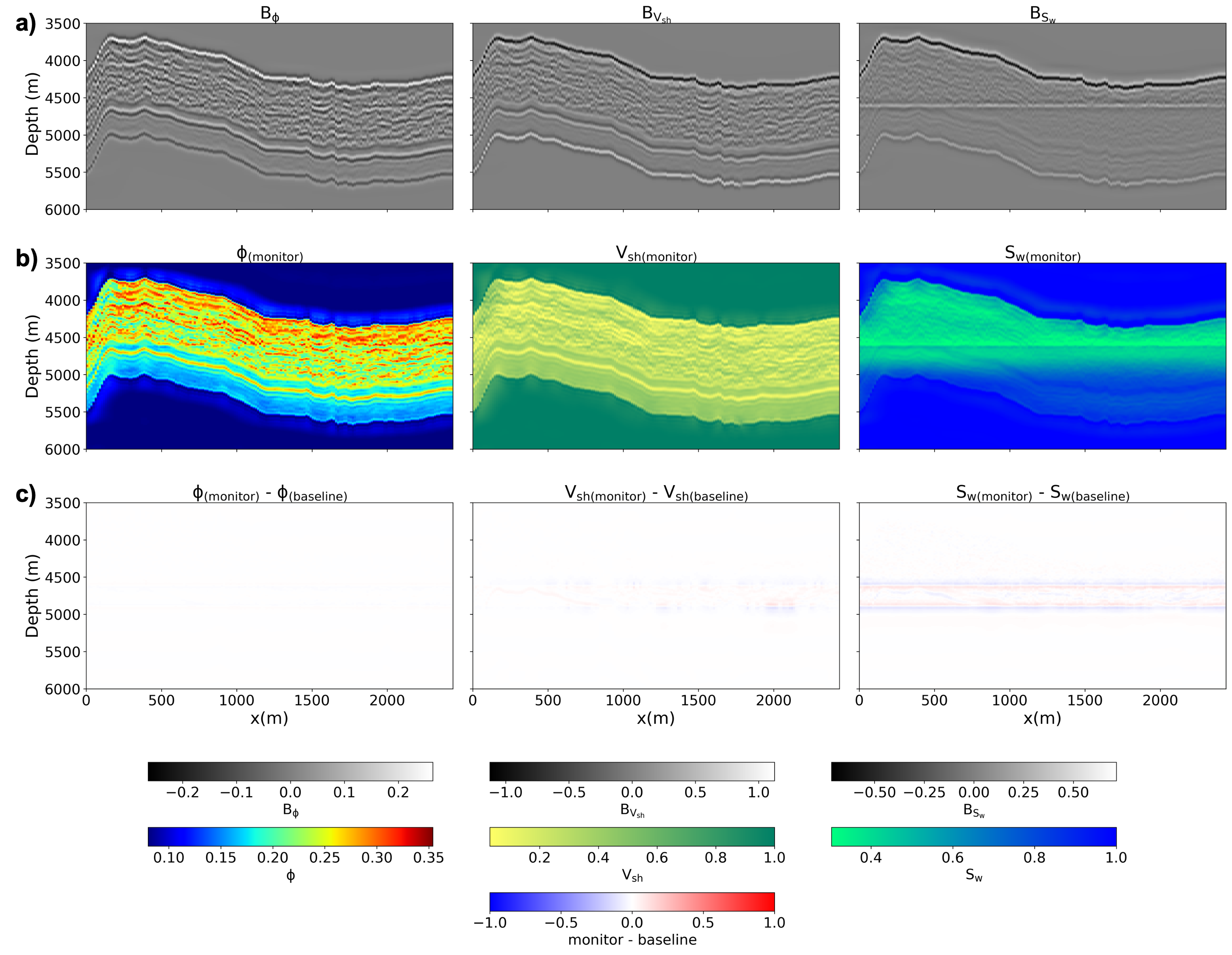}
    \caption{Inversion results for the model with updated oil-water contact. a) Petrophysical coefficients (B) obtained. b) Inversion results after water displacement (monitor). c) Difference between inverted properties before (baseline), and after (monitor) the  oil-water contact was displaced.}
    \label{fig:results_synthetic_4D_inversion}
\end{figure}

Next, we further investigated the capabilities of Seis2Rock in the context of geophysical monitoring using pre-stack time-lapse seismic data. A second dataset was created by shifting the oil-water contact to Depth=100. Note that the training dataset, and therefore the basis functions, remain unchanged. The inversion results presented in Figure \ref{fig:results_synthetic_4D_inversion} show that our method can produce almost the same porosity and shale content values as the baseline data, as well as a highly accurate estimation of the oil-water contact movement. \\

Figure \ref{fig:results_synthetic_4D_inversion}c shows the difference between the petrophysical inversion results before and after the water displacement has taken place. Seis2Rock is able to identify and recover such fluid displacement.\\


\newpage
\subsection{Field data}

\begin{figure}[!h]
    \centering
    \includegraphics[width=\textwidth]{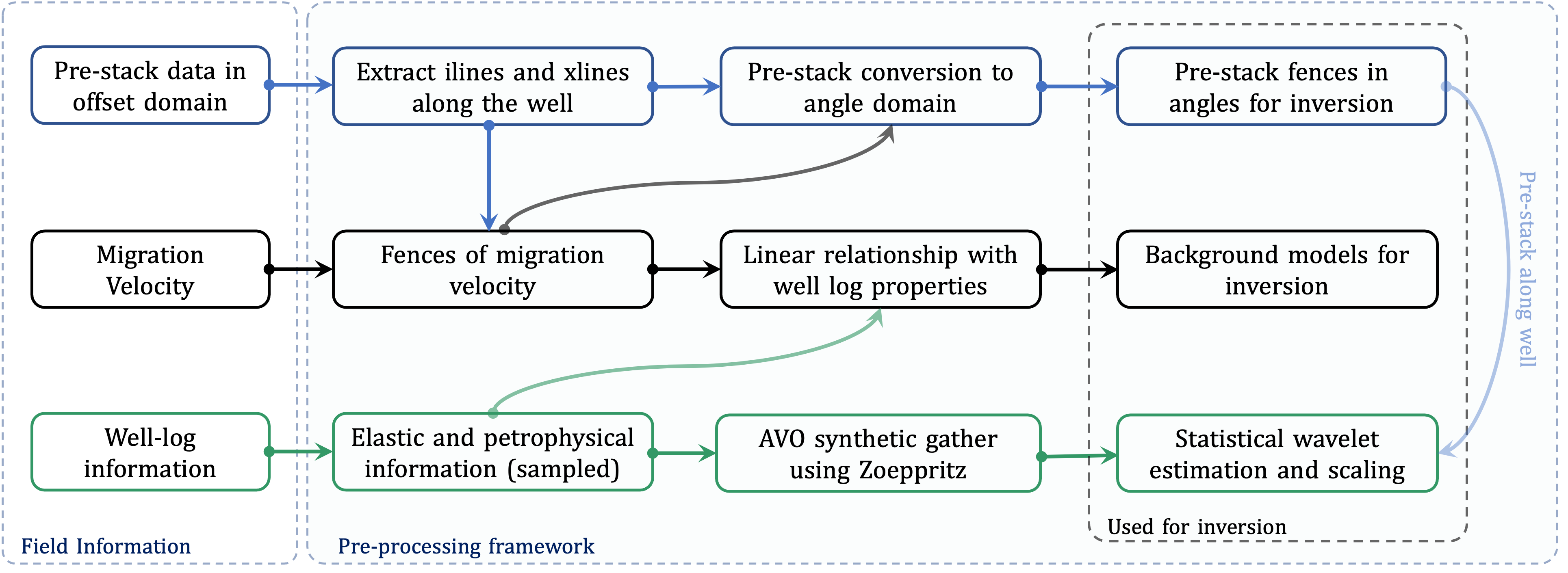}
    \caption{Summary of the pre-processing framework applied to the Volve dataset.}
    \label{fig:field_data_workflow}
\end{figure}

In order to apply Seis2Rock to a field dataset, one must have access to one or more wells with a well-log suite comprising of petrophysical and, ideally, elastic parameters, as well as time or depth pre-stack seismic offset (or preferably angle) gathers. The Volve dataset contains pre-stack seismic data in the offset domain and a vast collection of well logs with both petrophysical and elastic parameters from two wells. Therefore, Seis2Rock is employed here in a purely data-driven fashion, eliminating the need for a rock physics model. However, the application of the Seis2Rock methodology is not straightforward and a series of pre-processing steps is necessary to obtain accurate results. A summary of the pre-processing sequence adopted in this work is illustrated in Figure \ref{fig:field_data_workflow}.\\

\subsubsection{Pre-processing Volve data}
\begin{figure}[!h]
    \centering
    \includegraphics[width=0.75\textwidth]{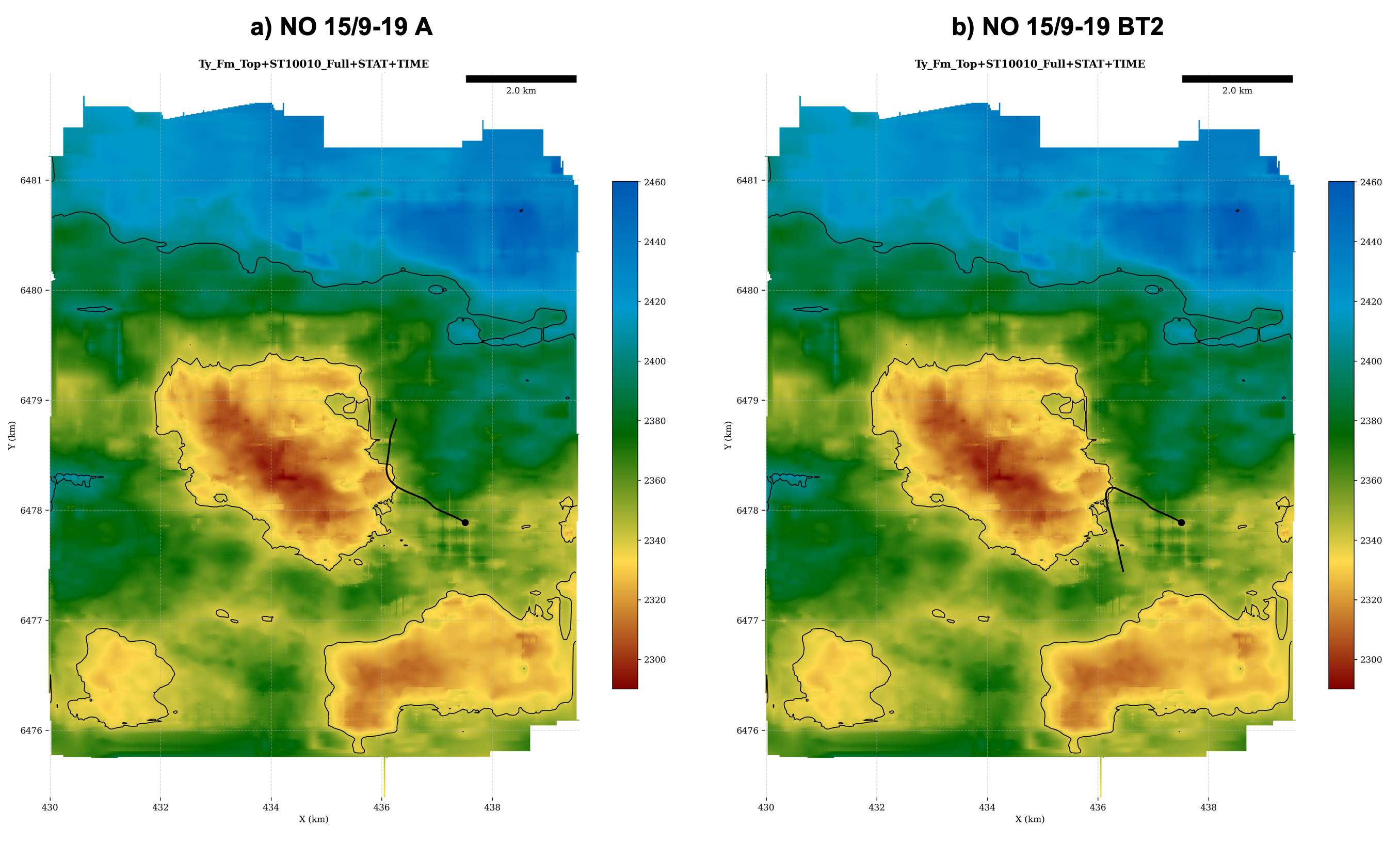}
    \caption{Well trajectories.}
    \label{fig:wells_trajectory}
\end{figure}

To begin with, we identify the three inputs required to apply the Seis2Rock workflow: a pre-stack seismic data in the offset domain, a migration velocity model, and well-logs containing elastic and petrophysical information. Wells NO 15/9-19 BT2 and NO 15/9-19 A provide appropriate data for our investigation. More specifically, wells NO 15/9-19 BT2 is a dry well (fully water saturated), whilst well NO 15/9-19 A is partially filled with oil. However, as their trajectories deviate (Figure \ref{fig:wells_trajectory}), we need to extract seismic data along 2D fences passing through the well paths. This is essential to enable the use of well-log information for comparison purposes after performing inversion.\\


\begin{figure}[!h]
    \centering
    \includegraphics[width=0.9\textwidth]{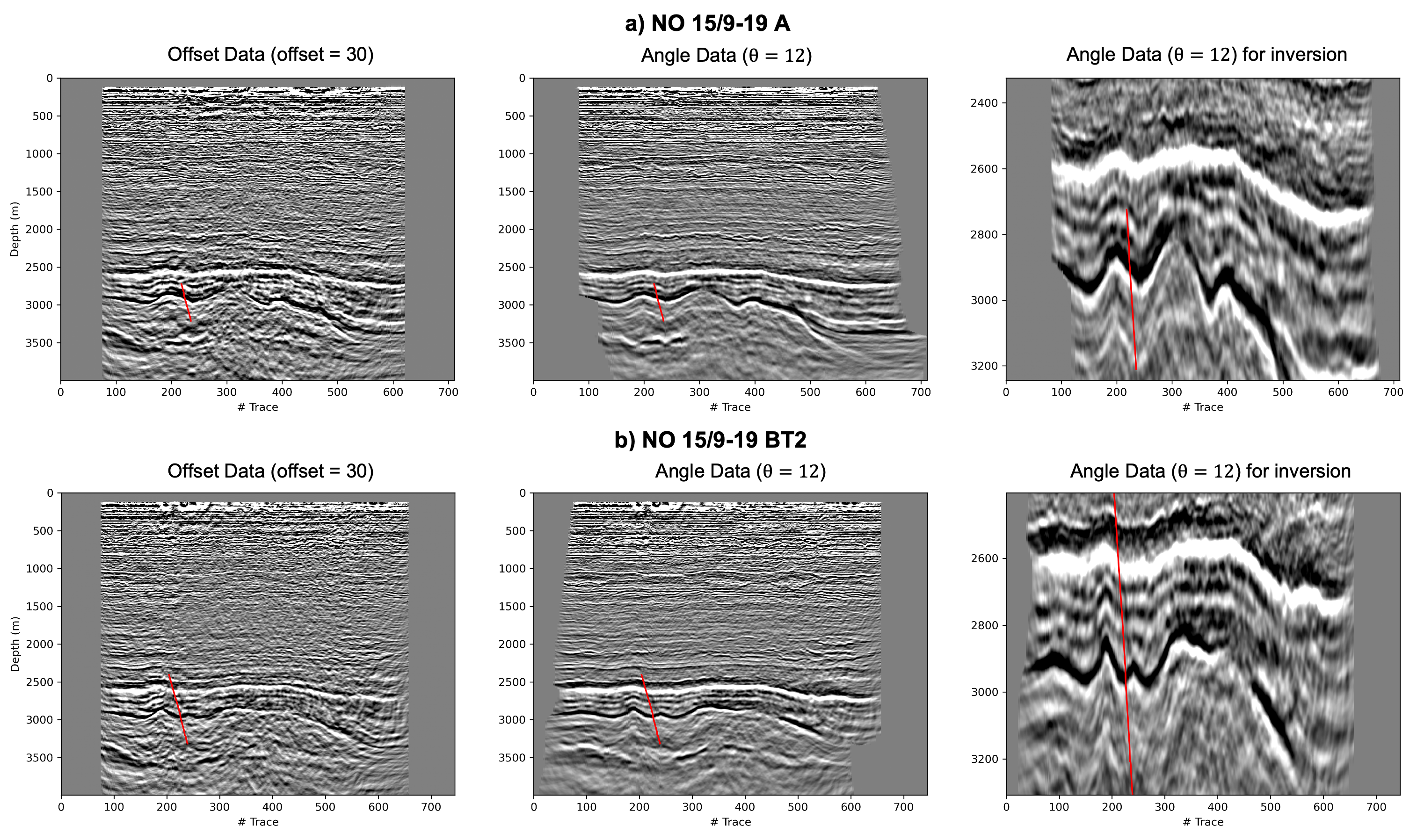}
    \caption{Fence data obtained along wells NO 15/9-19 A and NO 15/9-19 BT2. The offset domain, angle domain, and a subsection of the angle domain extracted for inversion are presented from left to right. The red line denotes where the well log information is collected.}
    \label{fig:fence_data}
\end{figure}

Given these field conditions, the pre-processing framework for our study, as illustrated in Figure \ref{fig:field_data_workflow}, begins with converting the pre-stack seismic data from the offset domain to the angle domain. We start the process by identifying the coordinates (ilines and xlines) corresponding to the deviated well paths. These coordinates serve as a basis for extracting the pre-stack data and migration velocity, limited only to the relevant ilines and xlines. The obtained pre-stack offset and velocity fences are both used to create the angle gathers. Figure \ref{fig:fence_data} presents the 2D fences along the two well logs used in this study. It shows the data in offset, its respective conversion in the angle domain, and the subsection to be used for inversion. \\

The next phase of our study entails the establishment of reliable background petrophysical models that serve as an initial guess for the final step of post-stack seismic inversion. We employ the petrophysical well-log information and velocity model at well locations to determine three different linear relationships. The resulting linear relationships are applied to the entire velocity model, which is converted into a set of petrophysical background models as required. \\

The last part of the pre-processing sequence is intended to create synthetic AVO gathers for the training phase on Seis2Rock. These synthetic AVO gathers should closely resemble those from the field data. To achieve this, we convert the well-logs to the same resolution as the pre-stack seismic data, followed by statistical wavelet extraction and amplitude calibration using the available pre-stack seismic gathers. As an example, the synthetic gather of well NO 15/9-19 BT2 is presented and compared to the real gather in Figure \ref{fig:field_inversion_well_1}. Though some differences can be observed between the two gathers, the key events are successfully modelled and present similar AVO responses to those in the field data. \\

\subsubsection{Seis2Rock inversion in Volve}
When applying Seis2Rock to the processed data, two primary steps are involved. The first step involves the inversion of only the gathers located at the well position to ascertain the method's validity. Subsequently, the methodology is extended to cover the pre-stack data extracted along the well fences.

\subsubsection*{Seis2Rock inversion along the well}
\begin{figure}[!h]
    \centering
    \includegraphics[width=\textwidth]{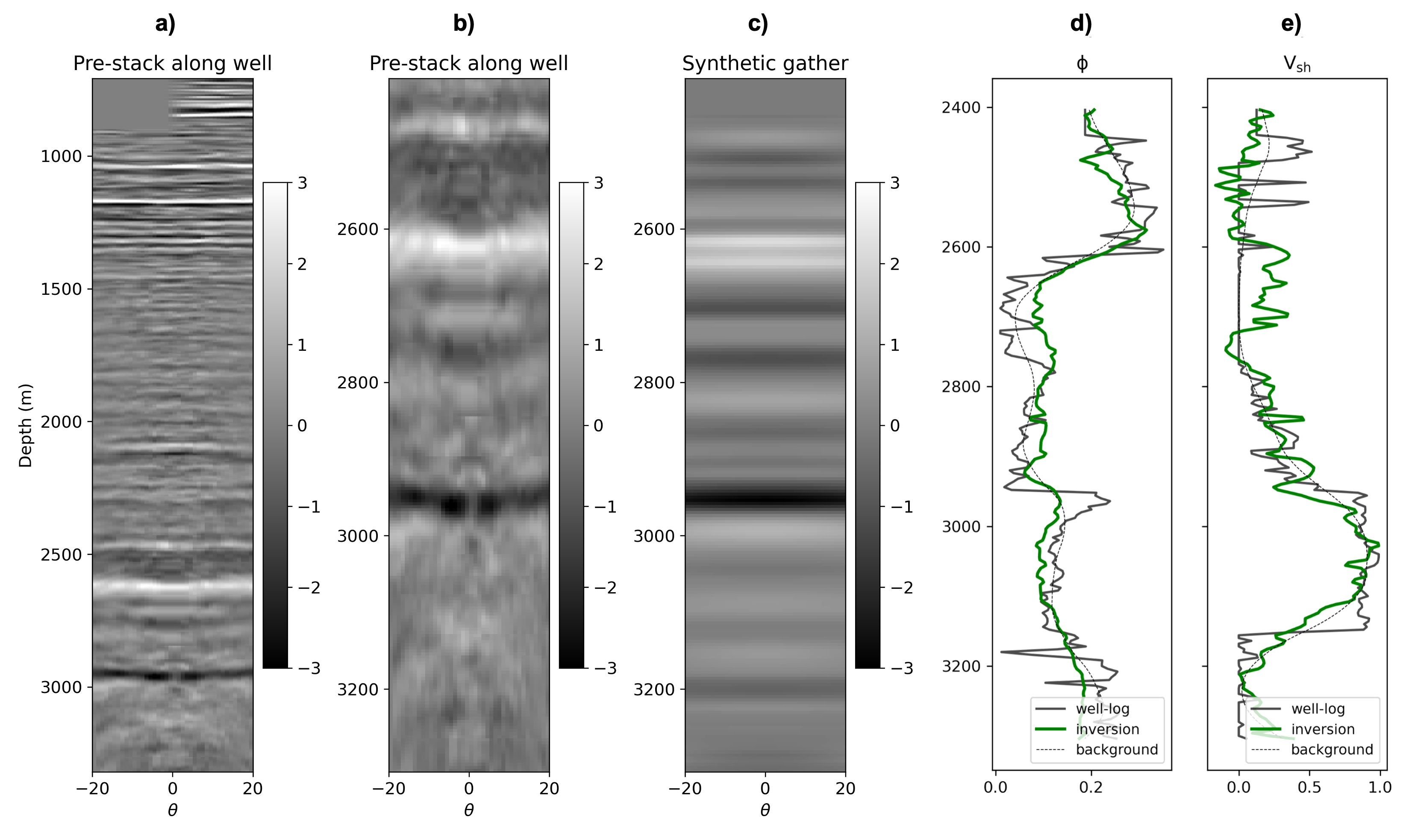}
    \caption{Seis2Rock inversion results along well NO 15/9-19 BT2 of the Volve field. a) Pre-stack data along the well trajectory. b) Close-up in a depth window where the well-log data is available. c) Synthetic gather created from the well-log data. d) Porosity ($\phi$) inversion results. e) Shale content ($
\mathrm{V_{sh}}$) inversion results. Water saturation is not inverted because the well-log presents a constant value for $\mathrm{S_w=1}$.}
    \label{fig:field_inversion_well_1}
\end{figure}

The accuracy of our method is first tested by extracting the pre-stack gather passing through well NO 15/9-19 BT2. The petrophysical and elastic well-log data are employed in the training stage to obtain the optimal basis functions and coefficients. We initially decided to work with a single gather to evaluate the ability of Seis2Rock to manage noise present in the data and assess whether we can reconstruct the petrophysical well logs used in training. The results of the Seis2Rock inversion are compared to the ground truth in Figure \ref{fig:field_inversion_well_1}d and e. Notably, the proposed methodology effectively reconstructs the well-log profile from a considerably smooth background model, with p=3 optimal basis functions. We opt for this number, as it successfully reduces the effect of noise in the data, whereas choosing larger values for the p coefficient produces poorer results. \\

\subsubsection*{Seis2Rock inversion along 2D fences}

\begin{figure}[h]
    \centering
    \includegraphics[width=\textwidth]{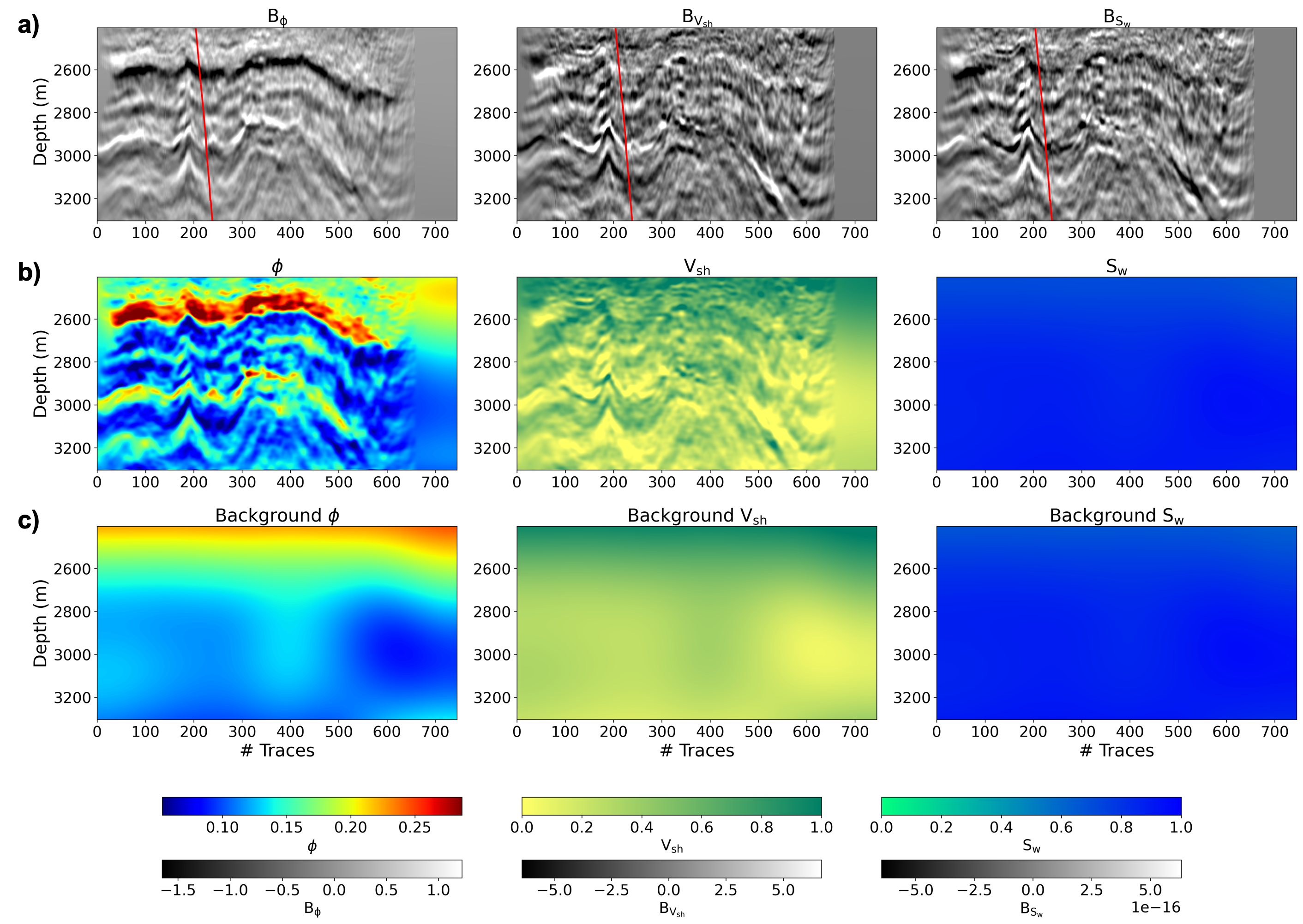}
    \caption{Seis2Rock inversion results on the fence along well NO 15/9-19 BT2 of the Volve field. a) Petrophysical coefficients data, with
the red line showing the well trajectory. b) Inversion results using only the information from well NO 15/9-19 BT2 in training. c) Background models used for inversion.}
    \label{fig:field_inversion_well_fence_1}
\end{figure}

\begin{figure}[h]
    \centering
    \includegraphics[width=\textwidth]{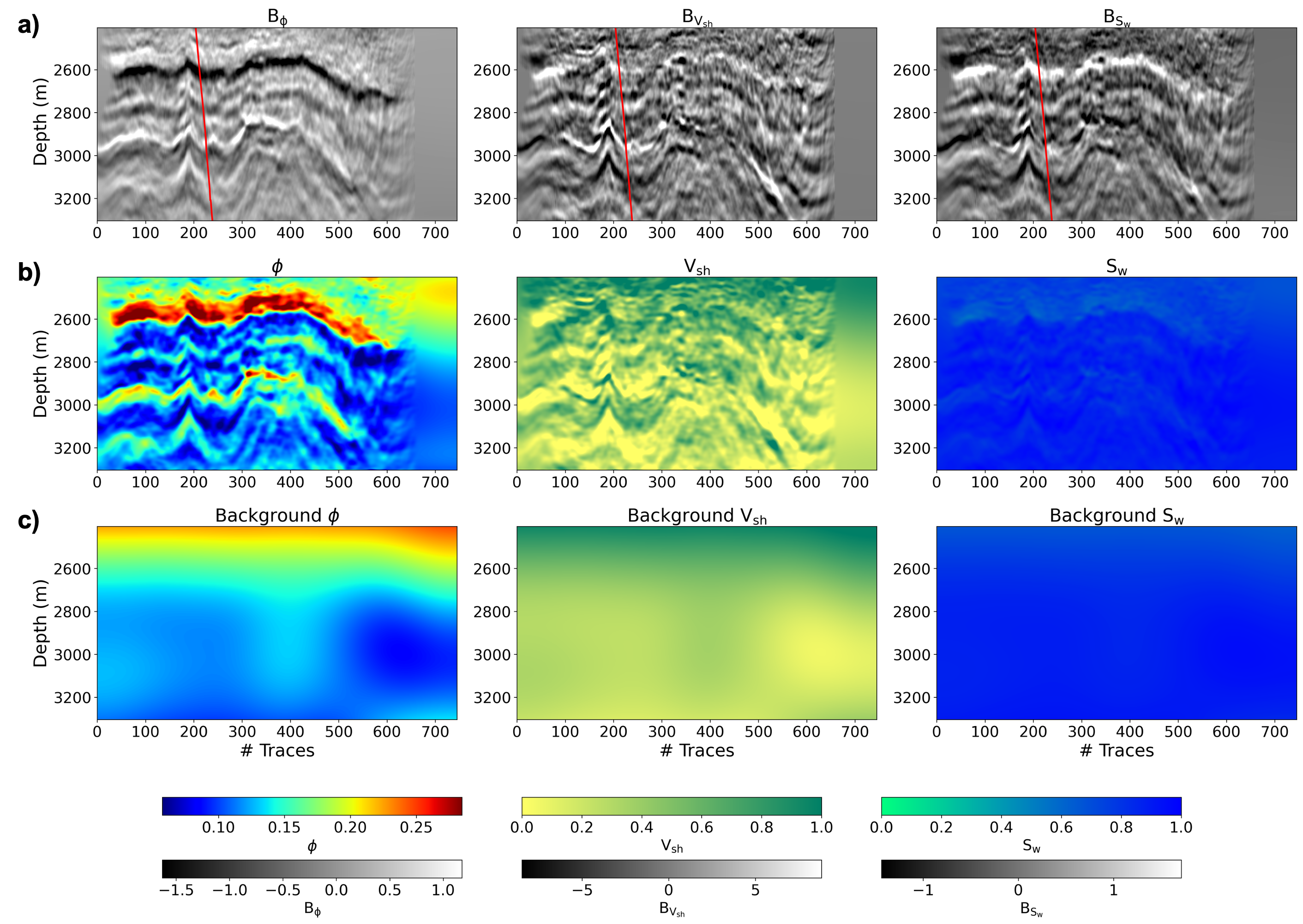}
    \caption{Seis2Rock inversion results on the fence along well NO 15/9-19 BT2 of the Volve field. a) Petrophysical coefficients data, with
the red line showing the well trajectory. b) Inversion results using both wells NO 15/9-19 BT2 and NO 15/9-19 A in training. c) Background models used for inversion.}
    \label{fig:field_inversion_well_fence_1_stacking}
\end{figure}

After successfully validating the ability of Seis2Rock to handle field data, we perform inversion on the pre-stack data along the two well fences extracted during the pre-processing stage. Initially, for the fence associated with well NO 15/9-19 BT2, we use the optimal basis functions computed in the previous section which utilize well-log data from the same well. Despite high noise levels within the data, Seis2Rock effectively reconstructs porosity and shale content along the fence. However, given the lack of hydrocarbon information within the saturation well-log (refer to Figure \ref{fig:field_inversion_well_fence_1}b), the inverted saturation model is almost identical to the background model. Subsequently, additional well-log information from well NO 15/9-19 A is incorporated in the training stage. Due to the presence of an oil zone in this well, the inverted water saturation model changes significantly, delineating potential hydrocarbon zones that were not identified in the previous inversion result (Figure \ref{fig:field_inversion_well_fence_1_stacking}b). However, some of these potential hydrocarbon zones may be erroneously identified due to the additional information provided by the second well in training. It is crucial to note that the outcomes for \ref{fig:field_inversion_well_fence_1}b) and \ref{fig:field_inversion_well_fence_1_stacking}b) were derived using the same background model, \ref{fig:field_inversion_well_fence_1}c). Thus, the hydrocarbon zones disclosed in \ref{fig:field_inversion_well_fence_1_stacking}b) are attributed to changes in the training data (due to the introduction of a second well), rather than a different background model causing water saturation changes.

\begin{figure}[!h]
    \centering
    \includegraphics[width=\textwidth]{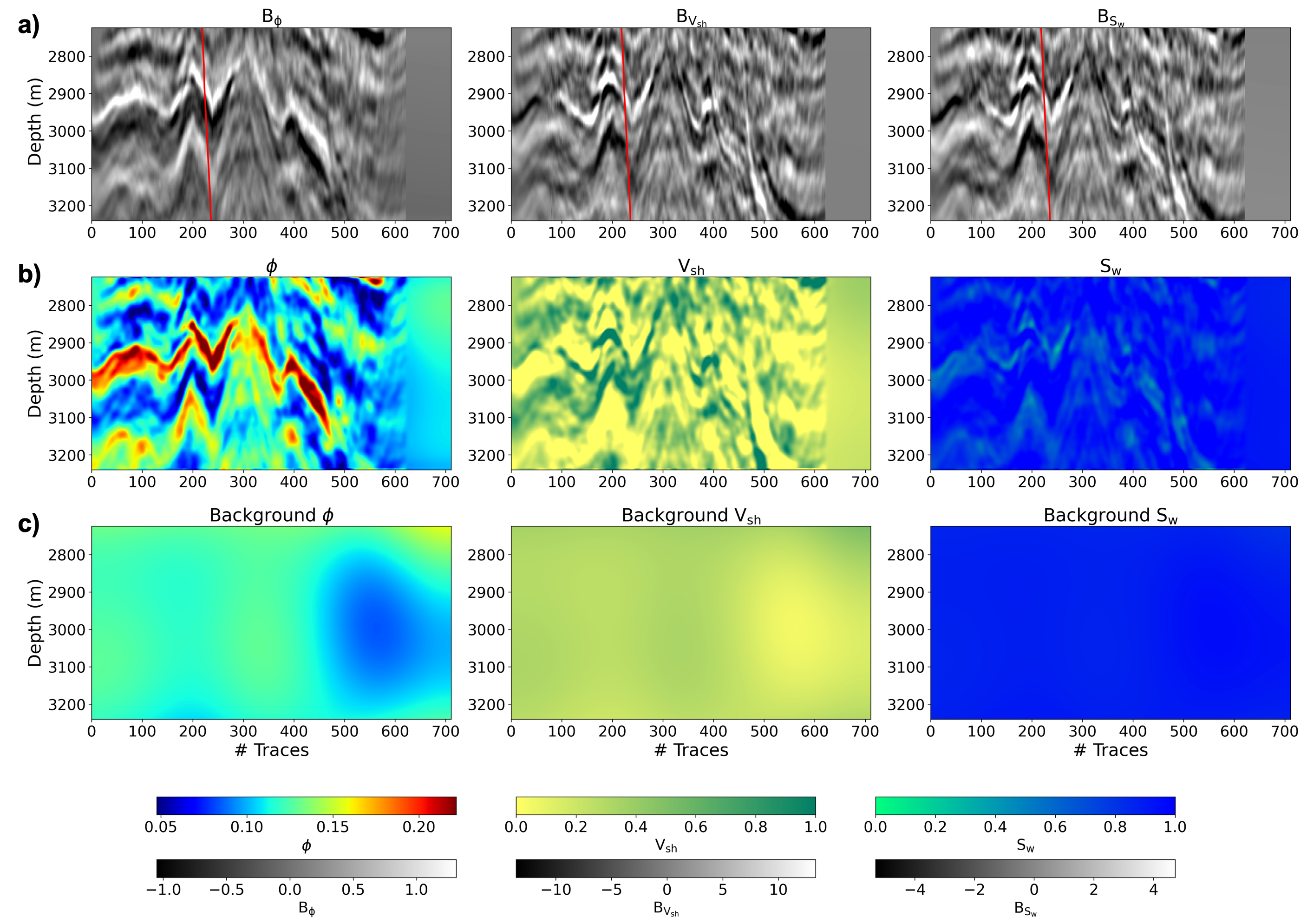}
    \caption{Seis2Rock inversion results for the 2D fence along well NO 15/9-19 A. a) Petrophysical coefficients data, with
the red line showing the well trajectory. b) Inversion results obtained using the well-log information from well NO 15/9-19 A in training. c) Background models used for inversion.}
    \label{fig:field_inversion_well_fence_0}
\end{figure}

\begin{figure}[!h]
    \centering
    \includegraphics[width=\textwidth]{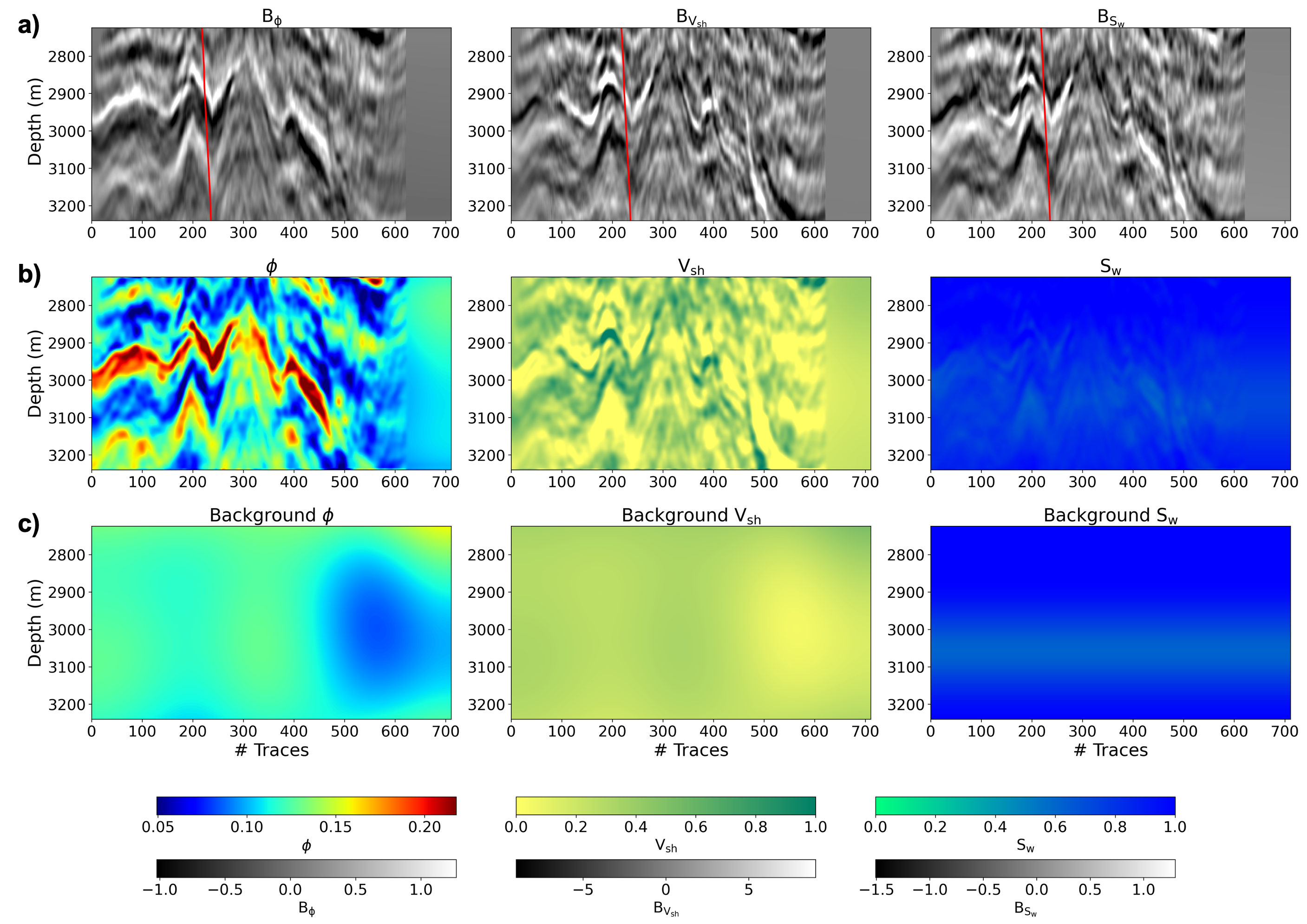}
    \caption{Seis2Rock inversion results for the 2D fence along well NO 15/9-19 A. a) Petrophysical coefficients data, with
the red line showing the well trajectory. b) Inversion results obtained using the well-log information from wells NO 15/9-19 A and NO 15/9-19 BT2 in training. c) Background models used for inversion. The construction of these models for inversion was obtained by utilizing a smoothed variant of the water saturation log, which effectively accounted for the presence of oil. This smoothed representation was duplicated across the two-dimensional fence to establish a robust background model.}
\label{fig:field_inversion_well_fence_0_stacking}
\end{figure}

 Next, for the fence associated with well NO 15/9-19, the optimal basis functions are computed using solely the information contained within this well, and also adding well NO 15/9-19 BT2. Figures \ref{fig:field_inversion_well_fence_0} and \ref{fig:field_inversion_well_fence_0_stacking} present the inversion results obtained for porosity, shale content, and water saturation, alongside their respective background models. The estimated models exhibit lateral continuity and areas of high porosity and presence of hydrocarbon content. \\

%


\clearpage
\subsubsection{Seis2Rock inversion in 3D}

\begin{figure}[!h]
    \centering
    \includegraphics[width=\textwidth]{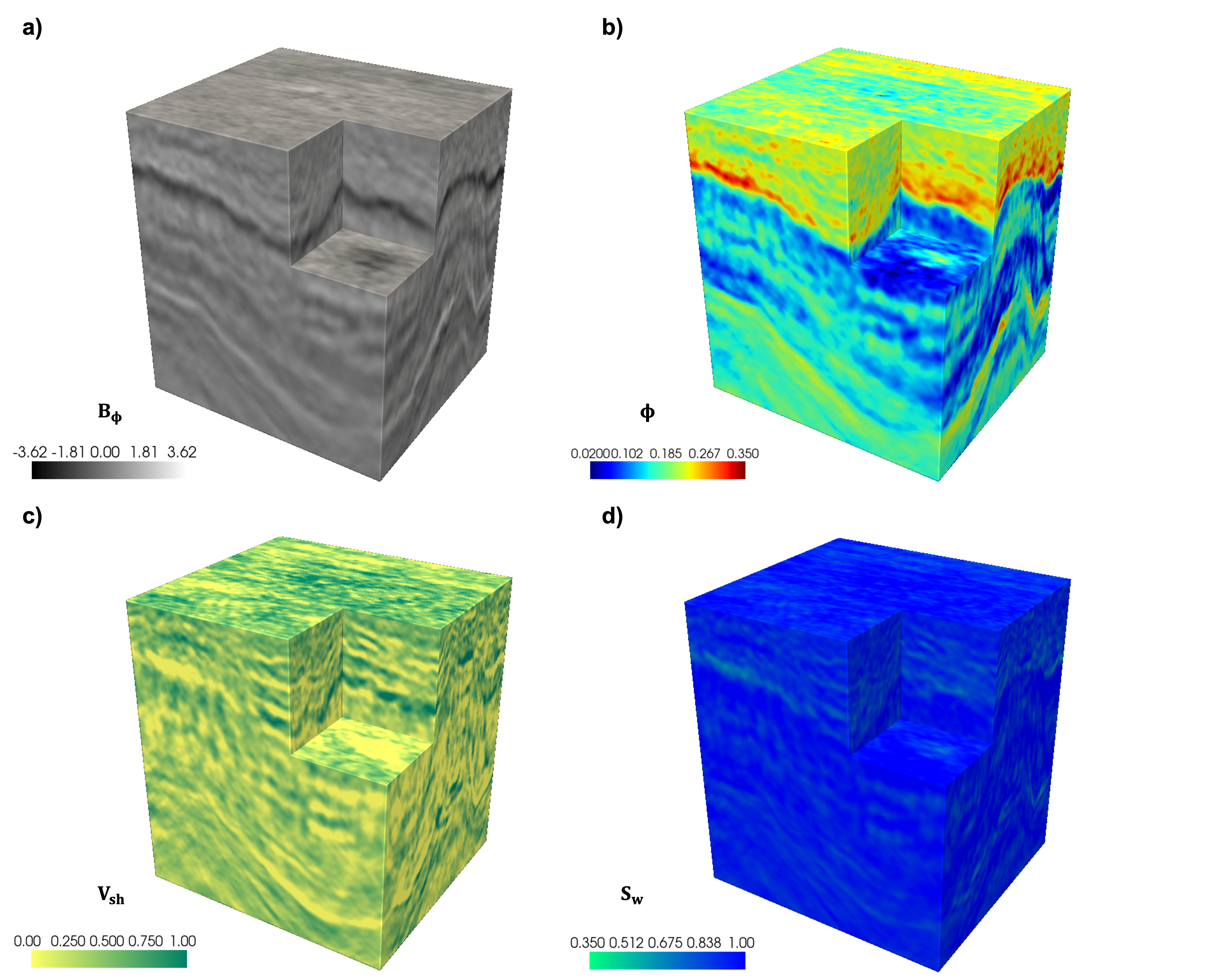}
    \caption{Seis2Rock inversion results for the 3D area of interest in the Volve dataset. a) Petrophysical coefficients B for porosity. Similar data terms are constructed for shale content and water saturation, however they are not included in the figure. Porosity, shale content, and water saturation results are show in b), c), and d) respectively.}
\label{fig:field_inversion_3D}
\end{figure}

In addition to our primary analyses, we also implemented a 3D inversion for each petrophysical parameter (porosity, shale content, and water saturation), thereby generating 3D petrophysical coefficient data for each specific parameter. For this case,  well-log information corresponding to wells NO 15/9-19 A and NO 15/9-19 BT2 was used to build the optimal basis functions in the Seis2Rock framework. The depth of the analysis area ranged from $2400 m$ to $3300 m$. In a similar vein, we utilized Laplacian regularization as a part of our inversion scheme. The results of each petrophysical parameter, along with the $\mathbf{B}$ coefficients tailored for porosity, are showcased in Figure \ref{fig:field_inversion_3D}.

The Seis2Rock inversion outcomes for the designated 3D area of interest are detailed as follows: part a) presents the petrophysical B coefficients specifically for porosity. Although we have also formulated similar data constructs for shale content and water saturation, these components are not depicted in the current figure. The results corresponding to the inversion of porosity, shale content, and water saturation are displayed in parts b), c), and d) respectively.

In conclusion, Seis2Rock accurately reconstructs petrophysical properties, particularly porosity and shale content, despite the high noise levels in the data. However, the water saturation displays inferior inversion results, potentially due to the limited variability of water saturation in the training data. \\

\clearpage
\section{Discussion}
Seis2Rock is a novel data-driven method for direct petrophysical inversion of pre-stack seismic data. A peculiar characteristic of the proposed method is that it relies on simple (linear) algebraic operations to invert an underlying nonlinear relationship (equation \ref{dgeneral}). Interestingly, the final step of the algorithm can be interpreted as a post-stack seismic inverse problem (although applied on so-called band-limited petrophysical parameters): 
being this a workhorse for quantitative characterization of the subsurface, many algorithms have been developed over the years, which we can directly benefit from. The retrieved petrophysical models are relatively smooth for both synthetic and field data, however, the high contrasts in petrophysical properties are underestimated in some regions. Whilst we attribute this behavior mostly to the choice of using a Laplacian regularization and least-squares solvers, future work will focus on addressing this limitation by employing alternative regularization and inversion techniques. Additionally, next research efforts could be directed towards imposing constraints on the iterative outcomes of the inversion process to ensure they remain within the specified bounds of the expected petrophysical values. \\

For example, Total Variation Regularization could be used to enhance blockiness in the recovered subsurface model. However, as applying TV regularization introduces a non-smooth functional in the loss function, this cannot be easily minimized with standard least-squares solvers. Proximal solvers, such as the alternating direction method of multipliers (ADMM)  can efficiently optimize these functionals \cite{wang2018data}. Moreover, integrating clustering or segmentation constraints into the inversion process, as exemplified in the joint inversion-segmentation approach of \cite{ravasi2022joint}, fosters the selection of models predominantly composed of a set of expected rock units or facies. In petrophysical inversion context, segmentation may be particularly beneficial as the link between facies and petrophysical properties is more direct than the link between facies and acoustic/elastic parameters as considered in \cite{ravasi2022joint}. In addition, our framework also permits the integration of new promising deep-based algorithms like the Plug-and-Play method with CNN-Based Denoisers \cite{romero2022plug} and its probabilistic extension \cite{izzatullah2022posterior, corrales2022bayesian}. \\

Apart from the final inversion step, Seis2Rock can be applied in an entirely data-driven fashion when well information includes both petrophysical and elastic parameters, as shown in the field data example. However, when elastic parameters are not measured at well locations, a representative rock physics model can be integrated to provide the link between petrophysical and elastic parameters; whilst this step allows Seis2Rock to be applied to a much wider set of use cases, it introduces additional complexity due to the uncertain nature of rock physical models. Future research will investigate how to embed sensitivity analysis and uncertainty quantification of the RPM and its hyperparameters into the Seis2Rock process. Finally, part of our method's robustness is attributed to its capacity to mitigate noise effects in seismic data by selecting an appropriate number of coefficients ($p$) when constructing the optimal basis functions and coefficients. However, the selection this parameter is user-dependent; consequently, incorporating $p$ into subsequent uncertainty quantification analysis is a viable approach. \\
\section{Conclusions}
Seis2Rock is an efficient and robust technique for petrophysical inversion. The introduced approach relies on singular value decomposition as a way to identify a set of optimal basis functions from pre-stack seismic data modelled at one or a small number of well locations. Such basis functions are later used to project pre-stack seismic data into band-limited petrophysical reflectivities; these reflectivities can be ultimately inverted for full-bandwidth petrophysical parameters by simply solving a post-stack seismic inversion per parameter. The proposed approach contrasts with data-hungry deep learning models, which require extensive amounts of synthetic data to establish a connection between petrophysical parameters and pre-stack data. Additionally, the flexibility to select the number of optimal coefficients allows Seis2Rock to handle data with various degrees of noise. Results on synthetic data indicate that Seis2Rock can directly invert petrophysical properties from seismic pre-stack data and porosity, water saturation, and shale content can be recovered with moderate to high degree of accuracy. When applied to field datasets, Seis2Rock relies on application of pre-processing steps to construct synthetic AVO gathers that closely mimic the field data.  The results obtained in this work on the Volve dataset suggest that Seis2Rock can effectively recover petrophysical information, even in the presence of high noise levels. In this regard, using a smaller number of optimal basis functions helps to manage noise levels. Finally, similarly to any other data-driven method, Seis2Rock's optimal basis functions may perform deficiently when applied on seismic datasets with geological settings that differ to those of the training data (e.g, far away from the control well).
\section*{Acknowledgment}
This publication is based on work supported by King Abdullah University of Science and Technology (KAUST) and the DeepWave consortium.  The authors also thank Equinor and partners for providing access to the \href{https://co2datashare.org/smeaheia-dataset/static/SMEAHEIA%20DATASET%20LICENSE_Gassnova%20and%20Equinor.pdf}{Smeaheia} and \href{https://www.equinor.com/energy/volve-data-sharing}{Volve} datasets. 

\newpage
\appendix
\section{\textsc{Seis2Rock back-projection of band limited optimal coefficients}}\label{appendix_A}

This Appendix provides a mathematical explanation of the back-projection process of Seis2Rock where optimal coefficients are converted into band-limited petrophysical parameters. To begin, we postulate the existence of a 3-terms linear modelling equation linking the petrophysical parameters of interest to pre-stack seismic data. We do so both for the data $\Tilde{d}(t_j, \theta)$ associated with the parameters coming from the available well-logs, as well as for the data $d(t_j, \theta)$ we wish to invert for:

\begin{equation}
    \label{avo_well_logs}
    \Tilde{d}(t_j, \theta) \approx  \sum_{i=1}^{N_t}  w(t_j-t_i) \left[\alpha \Tilde{r_1}(t_i) + \beta \Tilde{r_2}(t_i) + \gamma \Tilde{r_3}(t_i) \right]
\end{equation}

\begin{equation}
    \label{avo_seismic}
    d(t_j, \theta) \approx  \sum_{i=1}^{N_t}  w(t_j-t_i) \left[\alpha r_1(t_i) + \beta r_2(t_i) + \gamma r_3(t_i) \right]
\end{equation}

or in a matrix-vector notation as: 

\begin{equation}
    \label{avo_well_logs_mv}
    \mathbf{\Tilde{d}} = \alpha \mathbf{W} \mathbf{\Tilde{r_1}} + \beta \mathbf{W} \mathbf{\Tilde{r_2}} + \gamma \mathbf{W} \mathbf{\Tilde{r_3}}
\end{equation}

\begin{equation}
    \label{avo_seismic_mv}
    \mathbf{d} = \alpha \mathbf{W} \mathbf{r_1} + \beta \mathbf{W} \mathbf{r_2} + \gamma \mathbf{W} \mathbf{r_3}
\end{equation}

Here $\alpha$, $\beta$, $\gamma$ depend on the type of linearization, $r_1$, $r_2$, $r_3$ are reflectivity coefficients associated with the petrophysical parameters of interest (i.e., porosity, shale content, water saturation). $\mathbf{W}$ is the convolutional operator that applies the wavelet $w(t)$ to the reflectivities, and the symbol $\sim$ is used to indicate parameters coming from the well-log information. Note that in this derivation we have omitted for simplicity the background dataset.\\

We also expand the Seis2Rock modelling operator in equation \ref{gather} as:

\begin{equation}
    \label{d_svd}
    d_j(t_j, \theta) \approx c_1 f_1 (t_j, \theta) + c_2 f_2 (t_j, \theta) + c_3 f_3 (t_j, \theta) + \dots = \sum_{k=1}^p c_k f_k (t_j, \theta)
\end{equation}

From \cite{Causse2007} we can write each basis function $f$ as follows: 



\begin{equation}\label{basis_function_causse_2}
\begin{aligned}
f_k (t_j, \theta)      &= \sum_{j=1}^{N_t} h_{jk} \Tilde{d} (t_j, \theta)\\ 
    & = \sum_{j=1}^{N_t} h_{jk} \sum_{i=1}^{N_t} w(t_j-t_i) \left[ \alpha \Tilde{r_1}(t_i) + \beta \Tilde{r_2} (t_i) + \gamma \Tilde{r_3} (t_i) \right] \\
    & = \alpha \sum_{j=1}^{N_t} h_{jk} \sum_{i=1}^{N_t} w(t_j-t_i) \Tilde{r_1}(t_i) + \beta \sum_{j=1}^{N_t} h_{jk} \sum_{i=1}^{N_t} w(t_j-t_i) \Tilde{r_2}(t_i) + \gamma \sum_{j=1}^{N_t} h_{jk} \sum_{i=1}^{N_t} w(t_j-t_i) \Tilde{r_3}(t_i)  
\end{aligned}
\end{equation}

Now we can insert the basis functions in equation \ref{basis_function_causse_2} into equation \ref{d_svd}:

\begin{equation}
\label{d_rockavo_bf}
\begin{split}
d_j(\theta) \approx & \alpha \sum_{k=1}^{p} c_k \sum_{j=1}^{N} h_{jk} \sum_{i=1}^{N} w(t_j-t_i) \Tilde{r_1}(t_i) + \\ 
& \beta \sum_{k=1}^{p} c_k \sum_{j=1}^{N} h_{jk} \sum_{i=1}^{N} w(t_j-t_i) \Tilde{r_2}(t_i) + \\ 
& \gamma \sum_{k=1}^{p} c_k \sum_{j=1}^{N} h_{jk} \sum_{i=1}^{N} w(t_j-t_i) \Tilde{r_3}(t_i) 
\end{split}
\end{equation}

Expressing equation \ref{d_rockavo_bf} for all time samples in the matrix vector notation: 
\begin{equation}
    \label{d_rockavo_bf_mv}
    \mathbf{d} = \alpha \mathbf{c^T} \mathbf{H^T} \mathbf{W} \mathbf{\Tilde{r_1}} + \beta \mathbf{c^T} \mathbf{H^T} \mathbf{W} \mathbf{\Tilde{r_2}} + \gamma \mathbf{c^T} \mathbf{H^T} \mathbf{W} \mathbf{\Tilde{r_3}}
\end{equation}
and considering the terms with the same coefficients $\alpha$, $\beta$, $\gamma$ in equations \ref{avo_seismic_mv} and \ref{d_rockavo_bf_mv}, we obtain: 
\begin{equation}
    \label{reflectivity_1}
    \mathbf{W} \mathbf{r_i} = \mathbf{c^T} \mathbf{H^T} \mathbf{W} \mathbf{\Tilde{r_i}} ~~~~~ i=1,2,3 \mathrm{(number~of~physical~parameters)}
\end{equation}

This set of equations could be written in a compact form for the whole time sequence and three petrophysical parameters if we define $\mathbf{R} = [\mathbf{r_1}(t), \mathbf{r_2}(t), \mathbf{r_3}(t)]$ and $\mathbf{C} = [\mathbf{c_1}(t), \mathbf{c_2}(t), \mathbf{c_3}(t)]$: 
\begin{equation}
    \label{reflectivity_2}
    \mathbf{WR} = \mathbf{C^T} \mathbf{H^T} \mathbf{W} \mathbf{\Tilde{R}}
\end{equation}
Finally, to obtain reflectivities we can simply divide by the wavelet in each side: 
\begin{equation}
    \label{reflectivity_3}
    \mathbf{R} = \mathbf{W^{-1}} \mathbf{C^T} \mathbf{H^T} \mathbf{W} \mathbf{\Tilde{R}}
\end{equation}

Similarly, to obtain the petrophysical parameters the derivative operator can be also inverted for from $\mathbf{R}$ leading to equation \ref{m}.

\makenomenclature

\nomenclature{\(r\)}{physical coefficients or reflectivities}
\nomenclature{\(r_{\phi}\)}{Porosity reflectivity}
\nomenclature{\(r_{V_{sh}}\)}{Shale content reflectivity}
\nomenclature{\(r_{S_w}\)}{Water saturation reflectivity}
\nomenclature{\(\phi\)}{Porosity}
\nomenclature{\(V_{sh}\)}{Shale content}
\nomenclature{\(S_w\)}{Water saturation}
\nomenclature{\(K_{min}\)}{Bulk modulus of the mineral/rock}
\nomenclature{\(\mu_{min}\)}{Shear modulus of the mineral/rock}
\nomenclature{\(\mu_{sat}\)}{Shear modulus of the saturated rock}
\nomenclature{\(K_{dry}\)}{Effective bulk modulus}
\nomenclature{\(\mu_{dry}\)}{Effective shear modulus}
\nomenclature{\(P\)}{Pressure}
\nomenclature{\(C\)}{Coordination number}
\nomenclature{\(\nu_{min}\)}{Poisson ratio of the mineral/rock}
\nomenclature{\(K_{fl}\)}{Bulk modulus of the fluid mix}
\nomenclature{\(\rho_{fl}\)}{Density of the fluid mix}
\nomenclature{\(\rho_{min}\)}{Density of the mineral/rock}
\nomenclature{\(V_p\)}{Compressional wave velocity}
\nomenclature{\(V_s\)}{Shear wave velocity}
\nomenclature{\(g\)}{Non-linear rock-physics model of choice}
\nomenclature{\(\xi\)}{Set of hyper-parameters for the rock-physics modeling.}
\nomenclature{\(T\)}{Temperature}

\nomenclature{\(N_t\)}{Number of samples in the time/depth axis}
\nomenclature{\(N_{\theta}\)}{Number of angle gathers}
\nomenclature{\(\mathbf{\widetilde m}\)}{Chosen dictionary of petrophysical paramters}
\nomenclature{\(\mathbf{\widetilde m_b}\)}{Chosen dictionary of the background petrophysical paramters}
\nomenclature{\(\mathbf{\widetilde d}\)}{Modeled seismic gather of size ${N_{\theta} \times N_t}$}
\nomenclature{\(\mathbf{\widetilde d_b}\)}{Modeled background seismic gather of size ${N_{\theta} \times N_t}$}
\nomenclature{\(\mathbf{F}\)}{Matrix of eigenvectors of size $N_{\theta} \times N_t$}
\nomenclature{\(\mathbf{V}\)}{Matrix of eigenvectors of size $N_{\theta} \times N_t$}
\nomenclature{\(\boldsymbol\Lambda\)}{Matrix of Eigenvalues embedded in the diagonal of size $N_t \times N_t$}
\nomenclature{\(\mathbf{C}\)}{Matrix of optimal coefficients}
\nomenclature{\(p\)}{Subset of basis functions}
\nomenclature{\(\mathbf{C_p}\)}{Matrix of optimal coefficients taking $p$ basis functions}
\nomenclature{\(\mathbf{F_p}\)}{Matrix of eigenvectors (basis functions) taking $p$ basis functions}
\nomenclature{\(\mathbf{D}\)}{Pre-stack seismic gather}
\nomenclature{\(\mathbf{D_b}\)}{Pre-stack background synthetic seismic gather}
\nomenclature{\(\mathbf{B}\)}{Petrophysical parameters, one per petrophysical property}
\nomenclature{\(\mathbf{W}\)}{Wavelet operator}
\nomenclature{\(\mathbf{T}\)}{Time-derivative operator}
\nomenclature{\(\pmb{\widetilde{R}}\)}{Matrix of petrophysical reflectivities}
\nomenclature{\(n_m\)}{number of petrophysical parameters to invert (i.e., porosity, shale content, and water saturation.}

\printnomenclature
\newpage

\bibliographystyle{unsrt}  
\bibliography{references}

\end{document}